\documentclass[10pt,a4paper,superscriptaddress,aps,prd,nofootinbib,notitlepage]{revtex4-1}
\usepackage{lmodern}
\usepackage{verbatim}

\usepackage[T1]{fontenc}
\usepackage[utf8]{inputenc}
\usepackage{color}
\usepackage{amsmath}
\usepackage{amssymb}
\usepackage[]{subfigure}
\usepackage{graphicx}
\usepackage{natbib}
\usepackage{listings}
\usepackage{array}
\newcolumntype{P}[1]{>{\centering\arraybackslash}p{#1}}
\usepackage{esint}
\usepackage[unicode=true,
 bookmarks=true,bookmarksnumbered=false,bookmarksopen=false,
 breaklinks=false,pdfborder={0 0 1},backref=false,colorlinks=true]
 {hyperref}

\makeatletter

\definecolor{somegreen}{cmyk}{0,0.49,0.98,0.09}
\definecolor{red}{rgb}{1,0,0}
\definecolor{magenta}{cmyk}{0,1,0,0}
\definecolor{violet}{cmyk}{0,1,0,0}
\definecolor{darkgreen}{rgb}{0,0.65,0.05}
\definecolor{antiquefuchsia}{rgb}{0.33, 0.1, 0.89}

\makeatother

\begin{document}
\title{Fisher matrix for multiple tracers: all you can learn from large-scale structure without assuming a model}
\author{Renan Boschetti}
\email{renan.boschetti@usp.br}
\author{L. Raul Abramo}
\affiliation{Departamento de F\'{\i}sica Matem\'atica, Instituto de F\'{\i}sica, Universidade de S\~ao Paulo, Rua do Mat\~ao 
1371, CEP 05508-090, S\~ao Paulo, Brazil}
\author{Luca Amendola}
\affiliation{ITP, Ruprecht-Karls-Universität Heidelberg, Philosophenweg 16, 69120
Heidelberg, Germany}

\date{\today}

\begin{abstract}
The galaxy power spectrum is one of the central quantities in cosmology. It contains information about the primordial inflationary process, the matter clustering, the baryon-photon interaction, the effects of gravity, the galaxy-matter bias, the cosmic expansion, the peculiar velocity field, etc.. Most of this information is however difficult to extract without assuming a specific cosmological model, for instance $\Lambda$CDM and standard gravity. In this paper we explore instead how much information can be obtained  that is independent of the cosmological model, both at background and linear perturbation level. We determine the full set of model-independent statistics that can be constructed by combining two redshift bins and two distinct tracers. We focus in particular on the statistics $r(k,z_1,z_2)$, defined as the ratio of $f\sigma_8(z)$ at two redshift shells, and we show how to estimate it with a Fisher matrix approach. Finally, we forecast the constraints on $r$ that can be achieved by future galaxy surveys, and compare it with the standard single-tracer result. We find that $r$ can be measured with a precision from 3 to 11\%, depending on the survey. Using two tracers, we find improvements in the constraints up to a factor of two.
\end{abstract}

\maketitle

\section{Introduction}

The recent explosion of cosmological data 
has allowed to infer many crucial properties of our Universe. For instance, we now know the age of the Universe, its spatial curvature, the epoch at which acceleration begins, the level of clustering, etc., with a precision better than a few percent. However, in most cases, this knowledge actually depends on assuming a specific model, typically $\Lambda$CDM. If we change the underlying model, the statistical data analysis has to be redone, and the results will in general change. The answer to fundamental issues, as e.g. whether gravity is Einsteinian or not at large scales, will therefore depend on the specific cosmological model. Indeed, it is well known that constraints may vary depending on the assumed models -- see, e.g., \cite{2009JCAP...01..044M,2013MNRAS.436..854B,perico2019cosmic}.

An alternative and complementary approach  
is  to derive  measurements of quantities of
cosmological interest without first assuming a particular  model.
There are two ways in current literature in which this model-independent goal has been so far deployed. One consists in replacing $\Lambda$CDM with  mathematical parametrizations not tied to specific physical models -- see, e.g., \cite{Thomas:2011sf,2019ApJ...887...36L,2018JCAP...05..008C,2020arXiv200110887K}. 
The problem of this approach, however, is that one replaces  physically motivated  quantities with  phenomenological parameters or functions that have less direct or not univocal physical interpretation. 
The second approach, that we follow here, consists in measuring directly the physical quantities of interest by combining data in such a way to cancel out, whenever possible and to some extent, the dependence on the underlying cosmological model -- see also, e.g., \cite{Taddei:2014wqa,Taddei:2016iku}. 
In other words, the goal is to identify the maximal set of statistics  -- i.e., the combinations of real data that do not depend on theory parameters -- that have a direct physical meaning.

In a previous paper \cite{2019JCAP...06..030A} some of us discussed how to obtain such model-independent constraints on the redshift distortion parameter 
\begin{equation}
    \beta_{g}=\frac{f}{b_g} \; ,
\end{equation}
 where $f=d\log G / d\log a$ is the linear matter growth rate, $G$ is the growth function (i.e. the density contrast normalized to the present value of unity today), and $b_g$ is the galaxy-dark matter bias. All these quantities are generally both time- and scale-dependent. 
 
This paper is devoted to another such combination, namely:
\begin{equation}
    F(k,z) =f (k,z) \sigma_8(k,z) \; ,
\end{equation}
where $\sigma_8(k,z)=\sigma_8 \, G(k,z)$ is the mass variance at a radius of $8 \, h^{-1}$ Mpc and $\sigma_8$ is its value at the present epoch. This combination contains valuable information about the growth history of matter perturbations at the linear order and it has been receiving growing attention \cite{beutler20126df, howlett2015clustering, song2009reconstructing, reid2012clustering, samushia2014clustering, de2013vimos}. Arguably, this quantity can break the degeneracy among modified gravity models which predict the same expansion history. Therefore, the measurement of this quantity in a model-independent fashion can play a key role in cosmological tests of gravity with the upcoming data. 

In the Kaiser (linear) approximation, the galaxy power spectrum in redshift space, $P_g$ can be written in terms of $\beta$ and $F$ as a polynomial in the direction cosine $\mu$ of the angle between the line of sight and the Fourier wavevector $\vec k$ \cite{hamilton1998linear}:
\begin{eqnarray}
        P_g(k,z,\mu) &=& b_g^2(1+\beta_{g}\mu^2)^2 G^2 \sigma_8^2 P_0(k) \nonumber\\
      &=&(b_g^2 G^2 \sigma_8^2 + 2 b_g^2 \beta_g  G^2 \sigma_8^2 \mu^2 + F^2 \mu^4  ) P_0(k) \; ,
    \label{eq:Lin_Power}
\end{eqnarray}
where $P_0(k)$ expresses the shape of the power spectrum at the present time, normalized to $\sigma_8 (z=0)$. 
In the next section we will extend the power spectrum to mildly non-linear scales.

For every bin in $(k,z)$, one can fit the data for various values of $\mu$ and measure directly  the three coefficients of the $\mu$-polynomial in Eq. (\ref{eq:Lin_Power}), $A(k,z)=b_g^2 G^2\sigma_8^2 P_0(k), B(k, z)=b_g^2 \beta_{g}  G^2 \sigma_8^2 P_0(k)$ and $C(k, z)=F^2   P_0(k)$. The ratio of any two of the three coefficients  gives $\beta$, regardless of the
cosmological model and of the spectrum shape. 
To obtain $F$ in a similar model-independent way, however, one needs
to take the ratio of the third coefficient $C(k, z)$ at two different redshifts. One has then
\begin{equation}
\label{def:r}
    r^2(k,z_1,z_2)\equiv \frac{F^2(k,z_1)}{F^2(k,z_2)} \; .
\end{equation}
This is yet another observable that can be measured from the linear galaxy power spectrum without any assumption about cosmology, the bias, or the shape of the power spectrum. 
Notice that $F(k,z)$ {\it per se} cannot be measured directly, unless one specifies the value of $P_0(k)$.
The statistics $r(k,z_1,z_2)$ is the subject of this paper.

We  stress the fact that in this work we show how  to combine different redshift shells  into a model-independent observable. This is in contrast to the usual approach of adding information from different redshifts in the context of an assumed model (typically, $\Lambda$CDM), in order to improve cosmological constraints
 as is often done when forecasting the outcomes of cosmological surveys -- see, e.g. \cite{Fonseca:2015laa,2019JCAP...12..028F}.

The aim of being as model-independent as possible would be incomplete if we could not convert raw observables (angles and redshifts) into distances, and therefore into Fourier wavevectors. This requires the knowledge of the Hubble-Lema\^itre function $H(z)$ and of the angular-diameter distance $D_A(z)$ in the relevant redshift range. This can be obtained in various ways, as it has been shown in \cite{Amendola:2019lvy}. 
In particular, we point out that, since these distances are assumed to be already fixed, we do not include any additional information that could be gathered through the Alcock-Paczynski effect \cite{alcock1979evolution}. This not only considerably simplifies the treatment but also allows to distinguish the uncertainties due to the perturbations from those due to the background. A generalized Fisher matrix that includes simultaneously background and perturbation effects is of course also a possibility, but it would need to include the power spectrum of peculiar velocities, as shown in \cite{Amendola:2019lvy}. 
Hence, here we assume for simplicity that distances can be obtained in a model-independent way, and in this paper we focus exclusively on $r(k,z_1,z_2)$. In an analysis in which the uncertainties due to the background are included with priors, the constraints on $r$ would need to be marginalised over the uncertainties in $H(z)$ and $D(z)$. Therefore, the forecast presented here represents a lower limit of the true achievable precision and aims to highlight the advantages of the multi-tracer when combined with the model-independent approach. 


A second aim of this paper is to investigate the advantage of the multi-tracer technique applied to our statistics.
In any given survey, one might be able to identify two or more tracers of the large scale structure, i.e. extragalactic sources of different types (galaxies, quasars, Lyman-$\alpha$ systems, 21cm sources, X-ray sources, etc.). If those objects trace the same underlying density field, but with different biases, then the constraints on the clustering properties of the tracers with respect to the dark matter density can be significantly enhanced by considering the tracers separately, as first noted in \cite{Seljak:2009,McDonald:2009}, and as we have also shown specifically for the model-independent measurement of $\beta$ in \cite{2019JCAP...06..030A}. As we will show in this paper, we can also obtain improved, model-independent constraints on $r(k,z_1,z_2)$ by comparing tracers at different redshifts.

\section{Model independent constraints on $f\sigma_{8}(z)$ in the quasi-linear regime}

\subsection{One tracer, two redshift bins}

We shall employ two redshift bins in order to obtain
model-independent constraints on $F=f\sigma_{8}(z)$. 
Although the argument detailed above is exact in the linear regime, we know that even on relatively large scales there are corrections due to the non-linearities inherent to structure formation.
We consider here the first such correction in redshift space, which arises from the galaxy pairwise velocity dispersion (the Fingers-of-God effect). We also
allow implicitly for scale-dependent corrections to the matter growth rate, which may arise either from non-linear corrections, or from other effects like  free-streaming scales and growth in modified gravity theories.
Since we will confine our analysis to large scales, $k\le 0.1 h$ Mpc$^{-1}$, the non-linear velocity dispersion corrections will however play only a minor role.

In this regime, and using a simple model for the redshift distortion and the Fingers-of-God  (FoG) effect \cite{white2009forecasting, hamilton1998linear}, the matter power spectrum includes an extra factor with respect to Eq. (\ref{eq:Lin_Power}), and can be written as:
\begin{equation}
    P_{g}(k,z,\mu) = [b_{g}(k, z) + f(k, z)\mu^{2}]^{2}\sigma^{2}_{8}(k ,z)G^{2}_{FoG}(k, \mu, z)P_{0}(k) \; ,\label{eq:mnlp}
\end{equation}
 where $G_{FoG}(k, \mu, z) =  \exp{[-\mu^{2}k^{2}\sigma^{2}_{v}(z)/2]}$ is  the  small-scale smoothing   due to peculiar velocities \cite{peacock1994reconstructing, hamilton1998linear}. Here, $\sigma_{v}(z)$ is the pairwise velocity dispersion of tracers inside a halo in units of $h^{-1}$ Mpc. The parameter $\sigma_v(z)$ is an additional observable that we include in our analysis. We therefore take the non-linear velocity dispersions in each redshift slice to be free independent parameters, which is a quite conservative set up.

With the FoG effect, the ratio of the third coefficient of the expansion of $P_{g}(k,z)$ as a $\mu$-polynomial at  two different redshift slices now gives:
\begin{equation}
    r_{k}^{2}(k, \mu,z_{1}, z_{2}) 
    = r^{2}(k, z_{1}, z_{2}) \times \left[\frac{ G_{FoG}(k, \mu,z_{1})}{ G_{FoG}(k,\mu, z_{2})} \right]^{2}  \; 
    \label{Eq:r2}
\end{equation}
  i.e. Eq. (\ref{def:r}) times a  FoG factor. This factor will play a minor role on the scales we consider in this work. 

We will now derive an expression for the uncertainty with which we can measure the ratio $r$.
Our starting point is the Fisher matrix for a single tracer. Given a survey with a comoving volume $V$, we can estimate the (redshift-space) galaxy power spectrum over some  Fourier-space bin $\vec{k}$. Since redshift-space distortions preserve symmetry under rotations around the Fourier-space azimuthal angle $\Delta\phi_k$, the Fourier-space volume of a bin is given by $V_{k,\mu} = (1/2) 2 \pi k^2 \Delta k \Delta \mu / (2\pi)^3$, where the factor of $1/2$ comes from the reality condition for the density field.

Given a galaxy sample with mean number density $\bar{n}_g$ and a redshift-space spectrum $P_{g}(z,\vec{k})$,
it is convenient to introduce the adimensional quantity 
$\mathcal{P}_g (z; \vec{k}) = \bar{n} P_{g}(z,\vec{k})$. 
We assume that the distribution of the raw data, i.e. the power spectrum in each $z,k$-bin, is well approximated by
a Gaussian.
In terms of this variable the Fisher information matrix for the galaxy power spectrum is given by \cite{tegmark1997measuring, abramo2012full, carron2015information}:
\begin{equation}
    F [\ln {\cal{P}}] = {\cal{V}}_\mu \left( \frac{{\cal{P}}}{1+{\cal{P}}} \right )^2
\end{equation}
where ${\cal{V}}_\mu = V V_{k,\mu}$ is the phase space volume, which accounts for both the survey volume $V$ and the Fourier space volume corresponding to the bin $(k,\mu)$.
For convenience, we will also denote the Fisher matrix per unit of phase space volume as $\bar{F} = F / {\cal{V}}_\mu$. In this section we focus on the Fisher matrix $\bar F$, which is independent of the survey volume. In the next section we will explore the implications of our results for some specific future surveys.

Let us now consider the case when we have a single type of galaxy observed in two different redshift bins, so our set of variables  are $Y = \{ \ln \mathcal{P}(z_{1},\vec{k}), \ln \mathcal{P}(z_{2},\vec{k})\}$. If we assume that the redshift shells are independent, then the Fisher matrix for every choice of $z_1,z_2,k$ is simply:
\begin{equation}
\bar{F}[Y] =
\begin{pmatrix}
     \left[ \frac{\mathcal{P}_1}{1+\mathcal{P}_1}\right]^{2}  & 0 \\
     0 & \left[ \frac{\mathcal{P}_2}{1+\mathcal{P}_2}\right]^{2}
\end{pmatrix} \; ,
\label{Fisher_single}
\end{equation}
where $\mathcal{P}_{1,2} = \bar{n}(z_{1,2}) P_{g}(z_{1,2},\vec{k})$.
From now on, to make our notation more clear, we use barred indices to refer to redshift slices, e.g., $\mathcal{P}_{\bar{1}}$ refers to the slice $z_1$. Unbarred indices will refer to different tracers. 

In order to derive constraints on the ratio $r$, as defined in Eq. (\ref{def:r}), we project the Fisher matrix of Eq. (\ref{Fisher_single}) onto the set of new variables: 
\begin{equation}
X = \{ \log r, \log P_{\bar{1}},\log \beta_{\bar{1}}, \log \beta_{\bar{2}}, \log \sigma_{v\bar{1}}, \log \sigma_{v\bar{2}}\} \; ,\label{Eq:set1}
\end{equation}
where $P_{\bar{1}} \equiv
\bar{n}_{\bar{1}}b^{2}_{\bar{1}}\sigma^{2}_{8}(k,z_{\bar{1}})P_{0}(k)$ refers to the {\em real-space} spectrum in units of the number density -- i.e., {\em without} the $\mu$-dependence.
The set $X$ is the complete set of independent parameters that we can construct in the case of one tracer and two redshift bins -- indeed, we can write $P_{\bar{2}}$ in terms of the other parameters as:
\begin{equation}
    P_{\bar{2}} = q\frac{P_{\bar{1}}\beta_{\bar{1}}^{2}}{r^{2}\beta^{2}_{\bar{2}}} \; ,
    \label{Eq:P2rel}
\end{equation}
where $q\equiv n_{\bar{2}}/n_{\bar{1}}$ is the ratio of the number densities at each slice.  

We then can project the Fisher matrix (\ref{Fisher_single}) onto the set $X$ as:
\begin{equation}
    \bar{F}_\mu[X_{\sigma}, X_{\lambda}] = \sum_{\alpha , \beta= 1}^{2}\frac{\partial Y_{\alpha}}{\partial X_{\sigma}} \bar{F}[Y_{\alpha}, Y_{\beta}]\frac{\partial Y_{\beta}}{\partial X_{\lambda}} \;.
    \label{project_single}
\end{equation}
where the subscript  $\mu$ reminds us that the $\mu$-integration has still to be performed.
The resulting Fisher matrix will have zero determinant by construction, since we started from two observables (${\cal{P}}_{\bar{1}}$ and ${\cal{P}}_{\bar{2}}$), and made a variable change into a new set of five observables. However, we can obtain a non-singular Fisher matrix by adding the information (i.e., the Fisher matrices) for all the values of $\mu$. In the linear regime, this procedure is equivalent to obtaining constraints using the information from the multipoles $\ell = 0,2,4$ of the redshift-space power spectra (if we include the FoG effect, formally the sum should be over all even multipoles.)

Therefore, the $\mu$-average of $\bar F$ is:
\begin{equation}
    \bar{F}[X] = \frac12 \int_{-1}^{1} d\mu \, \bar{F}_\mu [X] \; ,
    \label{mu_ave}
\end{equation}
and its inverse is  a well-defined covariance matrix.

The  relative marginalized variances per phase-space unit are then:
\begin{equation}
    \sigma_{r}^2 = (\bar{F}^{-1})_{11}  \ \ \text{,} \ \ \sigma_{\beta_{\bar{1}}}^2 = (\bar{F}^{-1})_{33} \ \ \text{,} \ \  \sigma_{\beta_{\bar{2}}}^2 = (\bar{F}^{-1})_{55} \; ,
\end{equation}
 for $r$, $\beta_{\bar{1}}$, $\beta_{\bar{2}}$, whereas
\begin{equation}
    \sigma_{\sigma_{v\bar{1}}}^2 = (\bar{F}^{-1})_{44}
\end{equation}
is the marginalized relative error for $\sigma_{v}$.

\subsection{Two tracers, two redshift bins}

In order to discuss the advantages, if any, of the multi-tracer technique, we should agree on how to combine two tracers into a single survey.
If two tracers belonging to the same survey are combined to form one single survey (from now on denoted as {\it combined} survey), their number densities are simply added as:
\begin{eqnarray}
    n &=& n_{1} + n_{2} 
    \\ \nonumber
    &=& \bar{n}_{1}[1 + (b_{1} + f\mu^{2})G^{(1)}_{\rm FoG}P^{1/2}_{m}] + \bar{n}_{2}[1 + (b_{2} + f\mu^{2})G^{(2)}_{\rm FoG}P^{1/2}_{m}]  \; .
\end{eqnarray}
where $P_m^{1/2}=G\sigma_8 P^{1/2}_0(k)$, and an overbar denotes average quantities. We can now collect the terms and write:
\begin{equation}
    n = \bar{n}[1 + (b+ f\mu^{2})G_{\rm FoG}\delta_{m}] \; ,
\end{equation}
where $\bar{n} = \bar{n}_{1} + \bar{n}_{2}$, and the  bias of the combined tracer is:
\begin{equation}
    b \equiv  \frac{\bar{n}_{1}b_{1}G^{(1)}_{\rm FoG}+ \bar{n}_{2}b_{2}G^{(2)}_{\rm FoG}}{\bar{n}_{1}G^{(1)}_{\rm FoG} + \bar{n}_{2}G^{(2)}_{\rm FoG}} \; ,
    \label{b_eff}
\end{equation}
whereas the  FoG term for the combined tracer is:
\begin{equation}
    G_{\rm FoG} \equiv \frac{\bar{n}_{1}G^{(1)}_{\rm FoG} + \bar{n}_{2}G^{(2)}_{\rm FoG}}{\bar{n}_{1} + \bar{n}_{2}} \; .
    \label{g_eff}
\end{equation}
With the latter definition we can also define a combined dispersion velocity as:
\begin{equation}
    \sigma^2_{\rm eff}(z_{\bar{i}}) = -\frac{2}{k^{2}\mu^{2}}\ln \left[  \frac{\bar{n}_{1}(z_{\bar{i}})G^{(1)}_{\rm FoG} + \bar{n}_{2}(z_{\bar{i}})G^{(2)}_{\rm FoG}}{\bar{n}_{1}(z_{\bar{i}}) + \bar{n}_{2}(z_{\bar{i}})} \right] = -\frac{2}{k^{2}\mu^{2}}\ln \left[  \frac{G^{(1)}_{\rm FoG} + q_{\bar{i}}G^{(2)}_{\rm FoG}}{1 + q_{\bar{i}}} \right]  \; .
\end{equation}
where we defined $q_{\bar{i}} \equiv \bar{n}_{2}(z_{\bar{i}})/\bar{n}_{1}(z_{\bar{i}})$. Therefore, the FoG correction of the combined tracer will be:
\begin{equation}
    G_{\rm FoG} = \exp {\left[ -\frac{\mu^{2}k^{2}\sigma_{\rm eff}^{2}}{2}\right]}.
    \label{fog_eff}
\end{equation}

The combined redshift space distortion parameter follows from (\ref{b_eff}):
\begin{equation}
    \beta \equiv \frac{f}{b} =  \frac{(1 + gq)\beta_{1}\beta_{2}}{\beta_{2} + gq\beta_{1}} \; ,
    \label{beta_eff} 
\end{equation}
where $g \equiv \exp{[-\mu^{2}k^{2}(\sigma^{2}_{2} - \sigma^{2}_{1})/2]}$ and the subindex $\bar{i}$ is everywhere implicit. In the following, we will  compare two tracers versus a single-tracer combined survey according to the above formulae.

The multi-tracer Fisher matrix was first derived in \cite{abramo2012full,
abramo2013multitracer} -- see also \cite{2019JCAP...06..030A} for a derivation and notations closer to the ones we use in this paper.
For the two tracer case, we start from the multi-tracer Fisher matrix per unit of phase - space volume for the set $Y^{2t} = \{ \log \mathcal{P}_{\bar{1}1}, \log \mathcal{P}_{\bar{1}2}, \log \mathcal{P}_{\bar{2}1}, \log \mathcal{P}_{\bar{2}2}\}$, 
where we remind the reader that barred indices run over the redshifts, unbarred ones over the tracers. We have:
\begin{equation}
    \bar{F}[Y^{2t}] =\frac{1}{2}
    \begin{pmatrix}
         \frac{\mathcal{P}_{\bar{1}1}\mathcal{P}_{\bar{1}}}{(1+\mathcal{P}_{\bar{1}})} + \frac{\mathcal{P}^{2}_{\bar{1}1}(1-\mathcal{P}_{\bar{1}})}{(1+\mathcal{P}_{\bar{1}})^{2}} & \frac{\mathcal{P}_{\bar{1}1}\mathcal{P}_{\bar{1}2}(1-\mathcal{P}_{\bar{1}})}{(1+\mathcal{P}_{\bar{1}})^{2}} & 0 & 0 \\
         \frac{\mathcal{P}_{\bar{1}1}\mathcal{P}_{\bar{1}2}(1-\mathcal{P}_{\bar{1}})}{(1+\mathcal{P}_{\bar{1}})^{2}} & \frac{\mathcal{P}_{\bar{1}2}\mathcal{P}_{\bar{1}}}{(1+\mathcal{P}_{\bar{1}})} + \frac{\mathcal{P}^{2}_{\bar{1}2}(1-\mathcal{P}_{\bar{1}})}{(1+\mathcal{P}_{\bar{1}})^{2}} & 0 & 0 \\
         0 & 0 & \frac{\mathcal{P}_{\bar{2}1}\mathcal{P}_{\bar{2}}}{(1+\mathcal{P}_{\bar{2}})} + \frac{\mathcal{P}^{2}_{\bar{2}1}(1-\mathcal{P}_{\bar{2}})}{(1+\mathcal{P}_{\bar{2}})^{2}} & \frac{\mathcal{P}_{\bar{2}1}\mathcal{P}_{\bar{2}2}(1-\mathcal{P}_{\bar{2}})}{(1+\mathcal{P}_{\bar{2}})^{2}} \\
         0 & 0 & \frac{\mathcal{P}_{\bar{2}1}\mathcal{P}_{\bar{2}2}(1-\mathcal{P}_{\bar{2}})}{(1+\mathcal{P}_{\bar{2}})^{2}} & \frac{\mathcal{P}_{\bar{2}2}\mathcal{P}_{\bar{2}}}{(1+\mathcal{P}_{\bar{2}})} + \frac{\mathcal{P}^{2}_{\bar{2}2}(1-\mathcal{P}_{\bar{2}})}{(1+\mathcal{P}_{\bar{2}})^{2}}
    \end{pmatrix} \; ,
    \label{Fisher_2t}
\end{equation}
where  $\mathcal{P}_{\bar{i}\alpha} $ is the effective power of tracer $\alpha$ at the redshift bin $\bar{i}$, and $\mathcal{P}_{\bar{i}} = \sum_{\alpha}\mathcal{P}_{\bar{i}\alpha}$ is the total effective power of all tracers at the bin $\bar{i}$. As in the single tracer case, in contrast with the {\em redshift-space} clustering strength ${\cal{P}}_{\bar{i}\alpha}$, below we define the {\em real-space} clustering strength as $P_{\bar{i}\alpha} \equiv \bar{n}_{\bar{i}\alpha}b^{2}_{\bar{i}\alpha}\sigma^{2}_{8\bar{i}}P_{0}(k)$. 

The complete set of parameters we can construct in the case of two redshift bins and two tracers is:
\begin{equation}
X^{2t} = \{ \log r, \log P_{\bar{1}}, \log \beta_{\bar{1} 1}, \log \beta_{\bar{1}2}, \log \beta_{\bar{2} 1}, \log \beta_{\bar{2} 2}, \log \sigma_{\bar{1}1}, \log \sigma_{\bar{1}2}, \log \sigma_{\bar{2}1}, \log \sigma_{\bar{2}2} \} \; .
\label{X2t}
\end{equation}
This reduction is possible because of the following relations, which are extensions of the relation obtained in the case of a single tracer and two redshift slices, Eq. (\ref{Eq:P2rel}):
\begin{equation}
    P_{\bar{i}1} = \frac{P_{\bar{i}}}{Z_{\bar{i}}},\quad P_{\bar{i}2} = \frac{P_{\bar{i}}Y_{\bar{i}}}{Z_{\bar{i}}} \; ,
\end{equation}
where
\begin{equation}
    Z_{\bar{i}} =(1+q_{\bar{i}}) \left( \frac{1+gq_{\bar{i}}\frac{\beta_{\bar{i}1}}{\beta_{\bar{i}2}}}{1+gq_{\bar{i}}} \right)^{2}\text{,} \ \ \ Y_{\bar{i}} =q_{\bar{i}}\left( \frac{\beta_{\bar{i}1}}{\beta_{\bar{i}2}}\right)^{2} \; .
\end{equation}
The relations above, together with the relation between $P_{\bar{2}}$ and $P_{\bar{1}}$,
\begin{equation}
    P_{\bar{2}} = q_{1}\frac{Z_{\bar{2}}P_{\bar{1}}\beta^{2}_{\bar{1}1}}{Z_{\bar{1}}r^{2}\beta^{2}_{\bar{2}1}} \; ,
    \label{Eq:consP2}
\end{equation}
where $q_{1}\equiv n_{\bar{2}1}/n_{\bar{1}1}$, make it possible to reduce the system to the set $X^{2t}$. 
Finally, we project the Fisher matrix in Eq. (\ref{Fisher_2t})  onto the set $X^{2t}$ in Eq. (\ref{X2t}) :
\begin{equation}
    \bar{F}_\mu[X^{2t}_{\sigma}, X^{2t}_{\lambda}] = \sum_{\alpha, \beta = 1}^{4}\frac{\partial Y^{2t}_{\alpha}}{\partial X^{2t}_{\sigma}} \bar{F}[Y^{2t}_{\alpha}, Y^{2t}_{\beta}]\frac{\partial Y^{2t}_{\beta}}{\partial X^{2t}_{\lambda}} \; .
    \label{projection}
\end{equation}
The 8-parameter set $X^{2t}$ (for each $z$- and $k$-bin) represents the complete set of model-independent clustering observables that can be obtained in the mildly non-linear regime expressed by Eq. (\ref{eq:mnlp}). More observables can only be included by moving to higher-order correlators, or by analysing other observables such as lensing.

Now that we possess all the tools in place to study any two redshift bins and two tracers, we proceed to the evaluation of the Fisher matrices.

\subsection{RESULTS}
\label{results}

In this section we present the results 
 for the uncertainty in the ratio $r$, $\sigma_{r}$, using the information drawn from one (combined) tracer, and from two tracers. These results are independent of the survey, because the Fisher matrix per phase-space volume is independent of the phase space volume ${\cal{V}}$. We
will focus on the scale $k=0.1 \, h$ Mpc$^{-1}$, but,  since the dependence in $k$ is only in the velocity smoothing factors, the results for larger scales are almost unchanged: e.g., for  $k=0.01 \, h$ Mpc$^{-1}$ the uncertainty shifts by only $\sim$1 \%.
 Unless otherwise stated, we adopt the values 
 $\beta_{\bar{i}1} = 0.5$, $\beta_{\bar{i}2} = 1.0$, $\sigma^{1}_{\bar{i}v} = 4 h^{-1} $ Mpc, $\sigma^{2}_{\bar{i}v} = 2 h^{-1} $ Mpc, $r=1$ and  $q = q_{\bar{1}} = q_{\bar{2}}=q_{1}=1$,  where the superscripts in the dispersion velocities refer to the tracers. Now and hereafter, in order to guarantee that the Fisher matrices are well conditioned, we use a prior of $50\%$ for the pairwise dispersion velocities of each tracer, which means we sum a constant term of $1/\sigma
^{2}_{\ln \sigma_{v}} $ to the diagonal of the Fisher matrix corresponding to the velocity dispersion, where $\sigma_{\ln \sigma_{v}} = \sigma_{\sigma_{v}}/\sigma_{v} = 0.5$. These realistic (and quite conservative) priors serve to produce Fisher matrices which are better conditioned, however we should note that even without any priors we would obtain Fisher matrices with well defined inverses.
 We assume the same values at the two redshift slices for $\beta $ and $\sigma_{v} $ in order to better tell apart the gain of having two tracers from the gain of having two redshift slices. These values are realistic but otherwise arbitrary, and serve only the purpose of illustration.  
We make the numerical code publicly available, so the interested  reader can  explore these constraints further for other sets of parameters. \footnote{\textit{Mathematica} notebook at the link https://github.com/RenanBoschetti/Fisher\_constraints}

In all plots in this section, the solid lines refer to the case where we consider the two distinct tracers, while dot-dashed lines refer to the combined tracer with bias and FoG correction given, respectively, by equations $(\ref{b_eff})$ and $(\ref{g_eff})$.

\begin{figure}[t]
    \centering
    \includegraphics[width=0.6\linewidth]{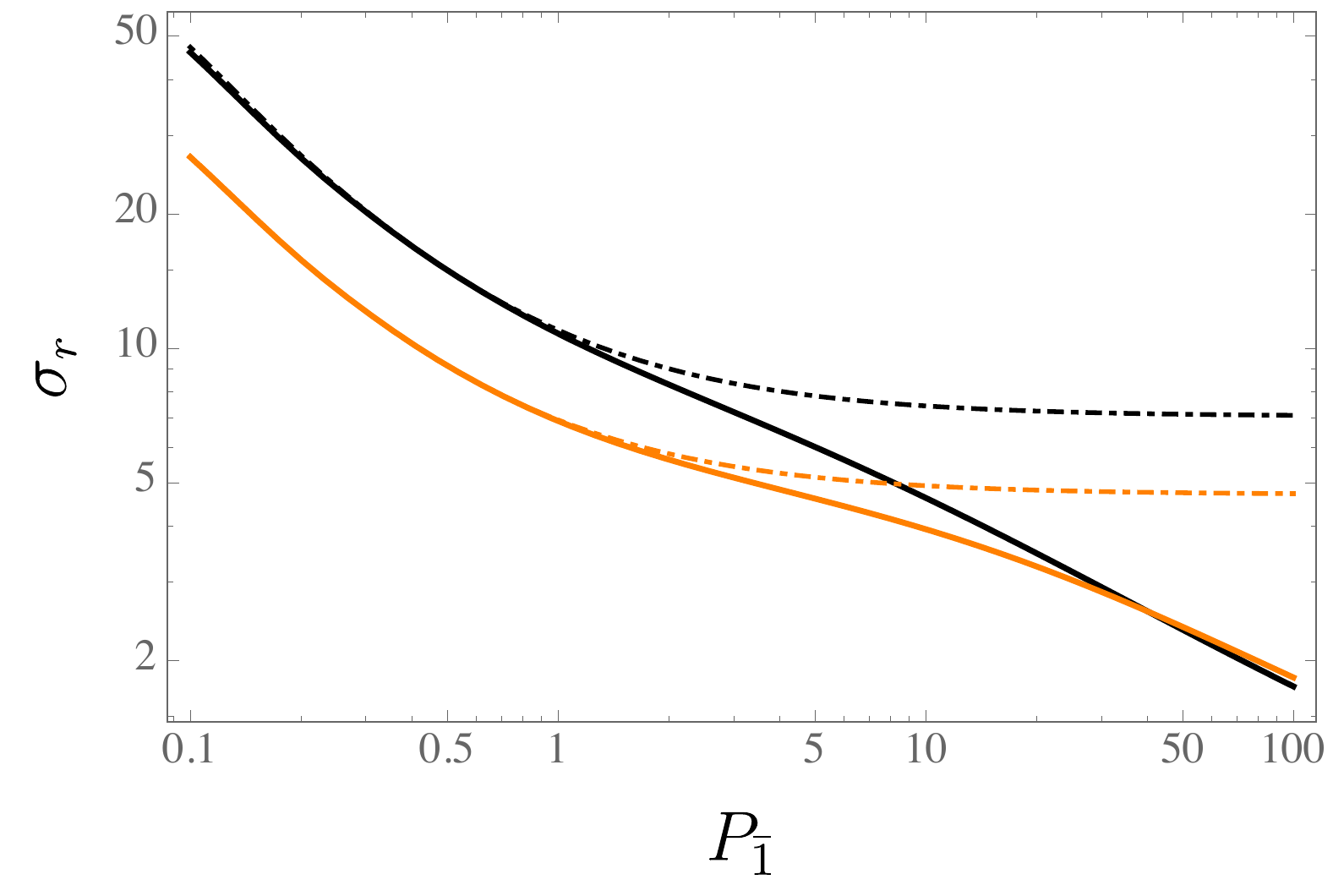}
    \caption{Marginalized relative errors $\sigma_{r}$ as a function of the real space clustering strength of the first redshift slice, $P_{\bar{1}}$, for the scale $k = 0.1 h$ Mpc$^{-1}$. Here, $\beta_{\bar{i}2} = 1.0$ is fixed. Black lines correspond to $\beta_{\bar{i}1} = 0.25$, and orange lines to $\beta_{\bar{i}1} = 0.5$. Solid lines refer to relative marginalized errors from the multi-tracer Fisher matrix, while the dot-dashed lines refer to the relative marginalized errors from the single-tracer Fisher matrix, for one combined tracer. As expected, the gain becomes noticeable only for $P\gg 1$.}
    \label{sigmar_p}
\end{figure}

Figure (\ref{sigmar_p}) shows the comparison between the relative marginalized error $\sigma_{r} = \sigma(r)/r$ in the cases of one combined tracer and two tracers, as a function of $P_{\bar{1}}$, for $\beta_{\bar{i}1} = 0.25$ (black) and $\beta_{\bar{i}1} = 0.5$ (orange). The error for the two tracers case always decreases as a function of the signal-to-noise ratio (SNR), $P_{\bar{1}}$, whereas the error for one tracer reaches a plateau. This is what we expect since the multi-tracer Fisher matrix is not limited by cosmic variance as its single-tracer counterpart. Therefore, for arbitrarily large SNR, the two-tracers errors are arbitrarily small.

\begin{figure}[t]
\centering
\includegraphics[width=0.6\linewidth]{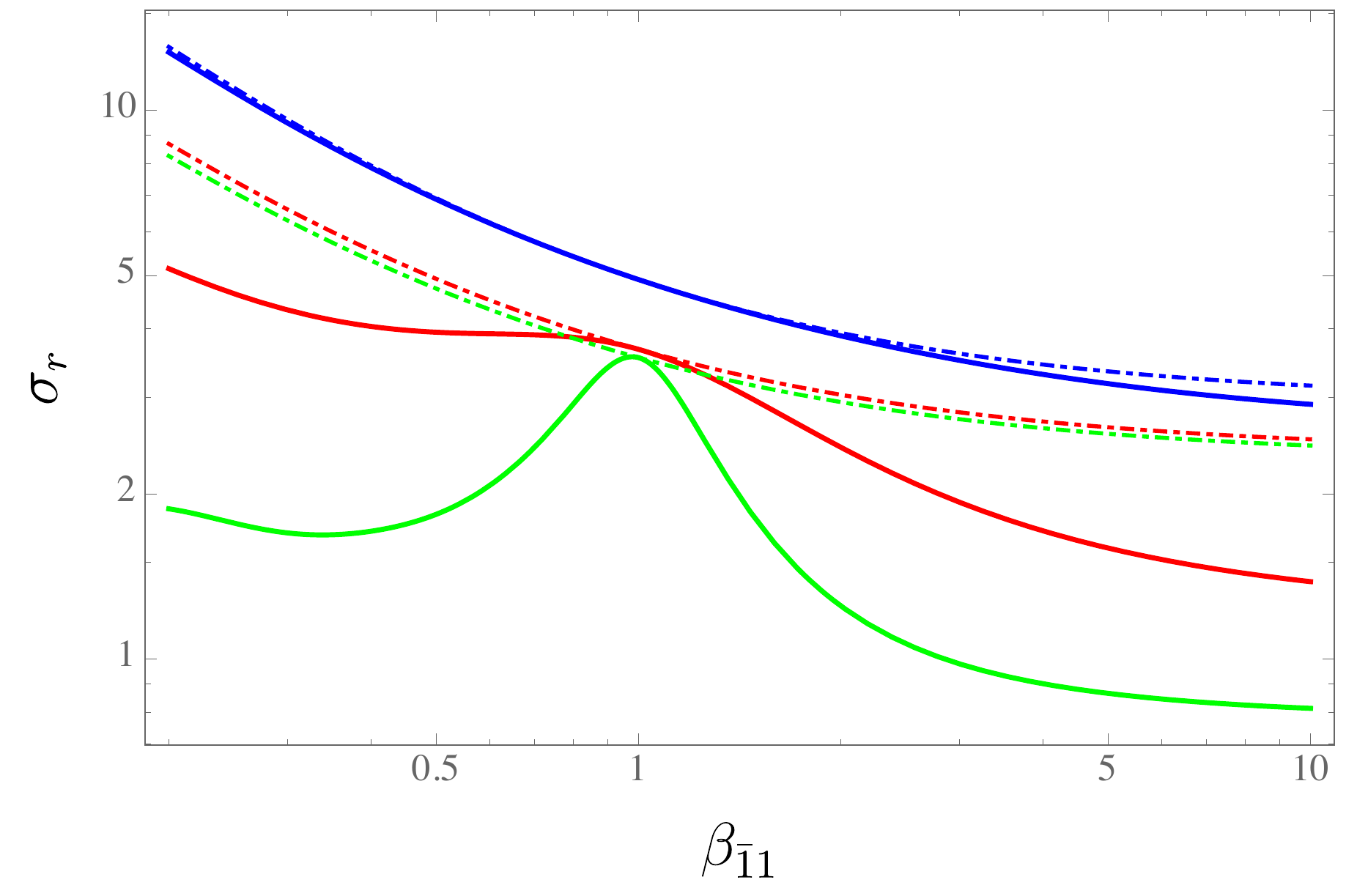}
\caption{Marginalized relative errors $\sigma_{r}$ as a function of $\beta_{\bar{1}1}$ for fixed $\beta_{\bar{i}2} = 1.0$ and $k = 0.1h$ Mpc$^{-1}$. Solid lines correspond to relative marginalized errors from the multi-tracer (two tracers) Fisher matrix while the dot-dashed lines correspond to the relative marginalized errors from the single-tracer Fisher matrix (combined tracer). The blue, red and green lines correspond respectively to $P_{\bar{1}} = 1$, $10$ and $100$. As expected, the combined tracer and two tracers constraints coincide at $\beta_{\bar{1}1}$, where the latter case reduces to the former.}
\label{sigmar_beta}
\end{figure}

Figure (\ref{sigmar_beta}) shows the relative marginalized error $\sigma_{r}$ as a function of $\beta_{\bar{i}1}$, while $\beta_{\bar{i}2} = 1.0$ is fixed, for $P_{\bar{1}} = 1.0$ (blue), $10.0$ (red) and $100.0$ (green). As expected, the single and two-tracers constraints coincide at $\beta_{\bar{i}1} = 1.0$, since in this case  the clusterings of the two tracers are in fact indistinguishable.

\begin{figure}[t]
\centering
\includegraphics[width=0.48\linewidth]{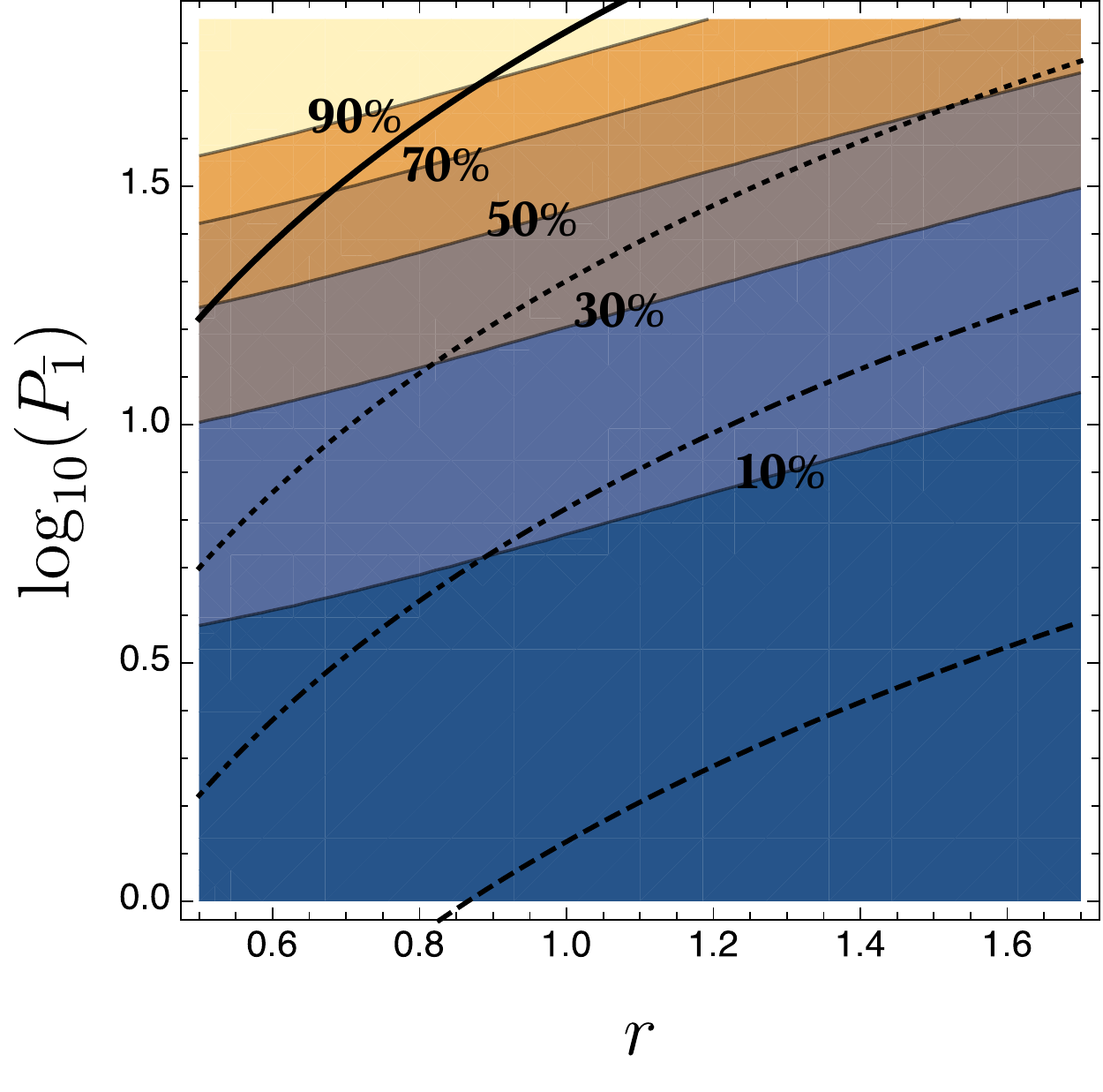}
\includegraphics[width=0.48\linewidth]{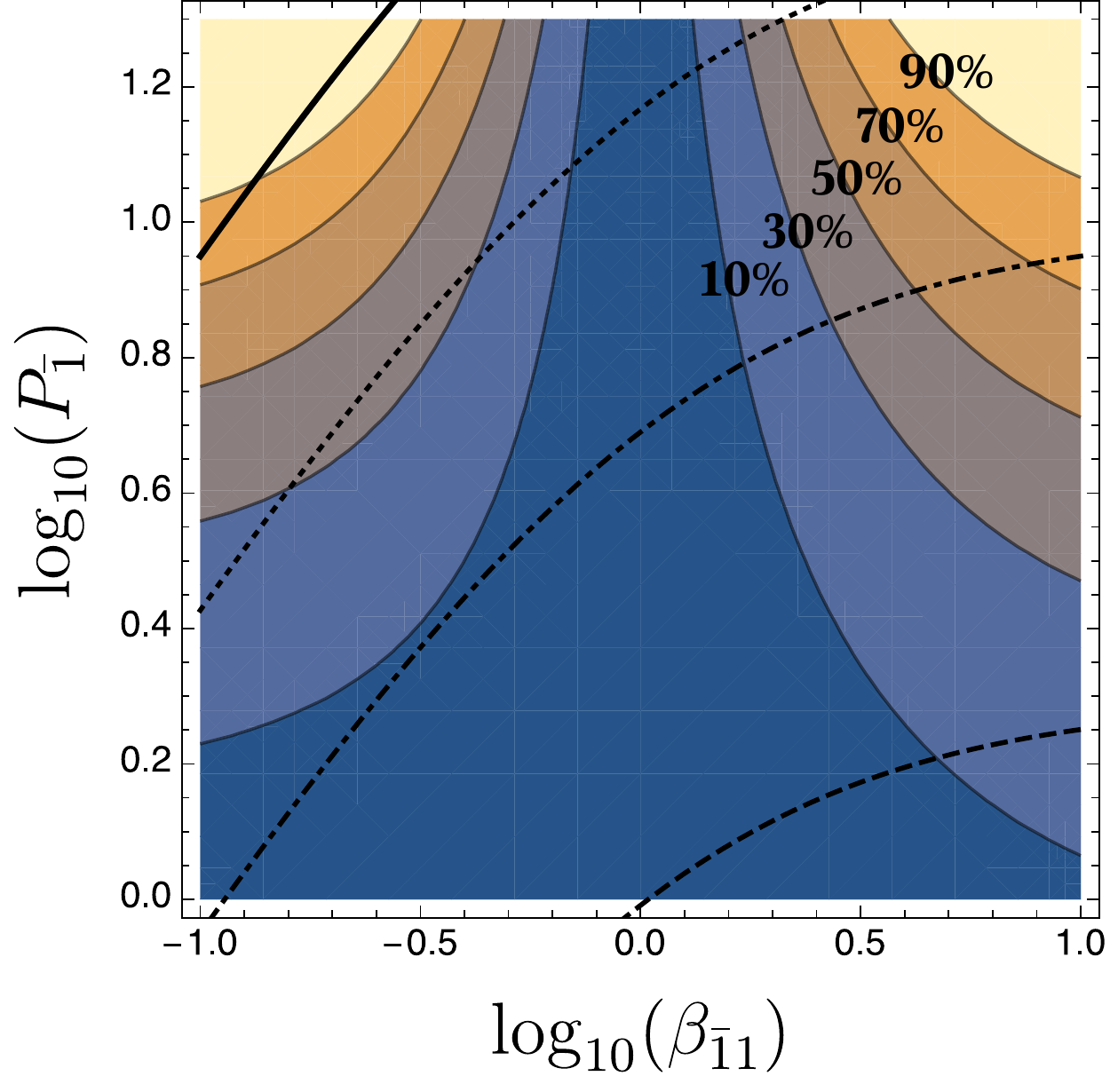}

\label{rel_diff}
\caption{Left: Relative difference $\Delta_{r} = \sigma^{1t}_{r}/\sigma^{2t}_{r}-1$ between relative marginalized errors $\sigma_{r}$ obtained with one-tracer Fisher matrix (combined tracer) and two-tracers Fisher matrix, represented by contours labelled by the percent gain (e.g, $10\%$ means $\Delta_{r} = 0.1$), in the $r-\log_{10}(P_{\bar{1}})$ plane. Right: the same relative difference in the $\log_{10}(\beta_{\bar{1}1})-\log_{10}(P_{\bar{1}})$ plane. Since $P_{\bar{2}}$ is constrained by Eqs. (\ref{Eq:consP2}), in both plots we denote the values
$P_{\bar{2}}=2.0$, $10.0$, $30.0$ and $100.0$, respectively, by the dashed, dot-dashed, dotted and solid lines. Comparing the left and right plots it is clear that the dependence of $\Delta_{r}$ on $r$ is much weaker than on $\beta_{\bar{1}1}$. The message here is that the value of $r$ plays a minor role  compared to $\beta_{\bar{1}1}$.}
\label{relative_diff}

\end{figure}

Finally, in Fig. (\ref{relative_diff}) we display the relative difference $\Delta_{r} \equiv \sigma^{1t}_{r}/\sigma^{2t}_{r}$ between constraints on $r$ from one tracer and two tracers in the $r$-$\log_{10}(P_{\bar{1}})$ (left) and $\log_{10}(\beta_{\bar{1}1})$-$\log_{10}(P_{\bar{1}})$ (right) planes along with curves of constant $P_{\bar{2}}$.   
As is already clear by the previous plots, the advantage of the two tracers approach increases with $P_{\bar{1}}$. This figure also shows the dependence of $\Delta_{r}$ on the SNR at the second redshift slice, $P_{\bar{2}}$: in the $r$-$\log(P_{\bar{1}})$ plane, the value of $P_{\bar{2}}$ drives the increase of $\Delta_{r}$. Indeed, it is clear that the dependence of $\Delta_{r}$ is much weaker on $r$ than it is on $\beta_{\bar{1}1}$. Futhermore, it is irrelevant whether we define $r$ or $1/r$: by swapping the slices, we achieve the same constraints.

\section{Constraining models}
\label{surveys_section}

\subsection{Fiducial model and surveys specifications}
In this section we apply the formalism to three realistic future surveys, namely Euclid \cite{laureijs2011euclid, amendola2018cosmology}, J-PAS (Javalambre Physics of the Accelerating Universe Astrophysical Survey) \cite{benitez2014j} and DESI (Dark Energy Spectroscopic Instrument) \cite{aghamousa2016desi, vargas2019unraveling}. The Euclid survey is a space telescope  that will map 15000 deg$^{2}$ of the sky. The J-PAS survey is a ground telescope, which aims to map 8500 deg$^{2}$ of the sky and will have the first light in 2020. The DESI survey is a ground telescope which will map 14000 deg$^{2}$ of the sky and should be fully operative already by 2021. 
At the same time, we show how to employ the general model-independent results of the previous section to constrain specific parametrizations. 

In order to derive the constraints in this section, we assume the flat $\Lambda$CDM as fiducial model, with cosmological parameters $\Omega_{c}h^{2} = 0.12$, $\Omega_{b}h^{2} = 0.022$, $h = 0.6732$ and $\sigma_{8}(z = 0) = 0.81$, which correspond to Planck 2018 \cite{aghanim2018planck} parameters. It should be stressed that despite assuming a model to derive these constraints, the method we propose remains model-independent. 
Once we have real data, in fact, they will replace our fiducial cosmology.

\begin{figure}
    \centering
    \includegraphics[scale = 0.5]{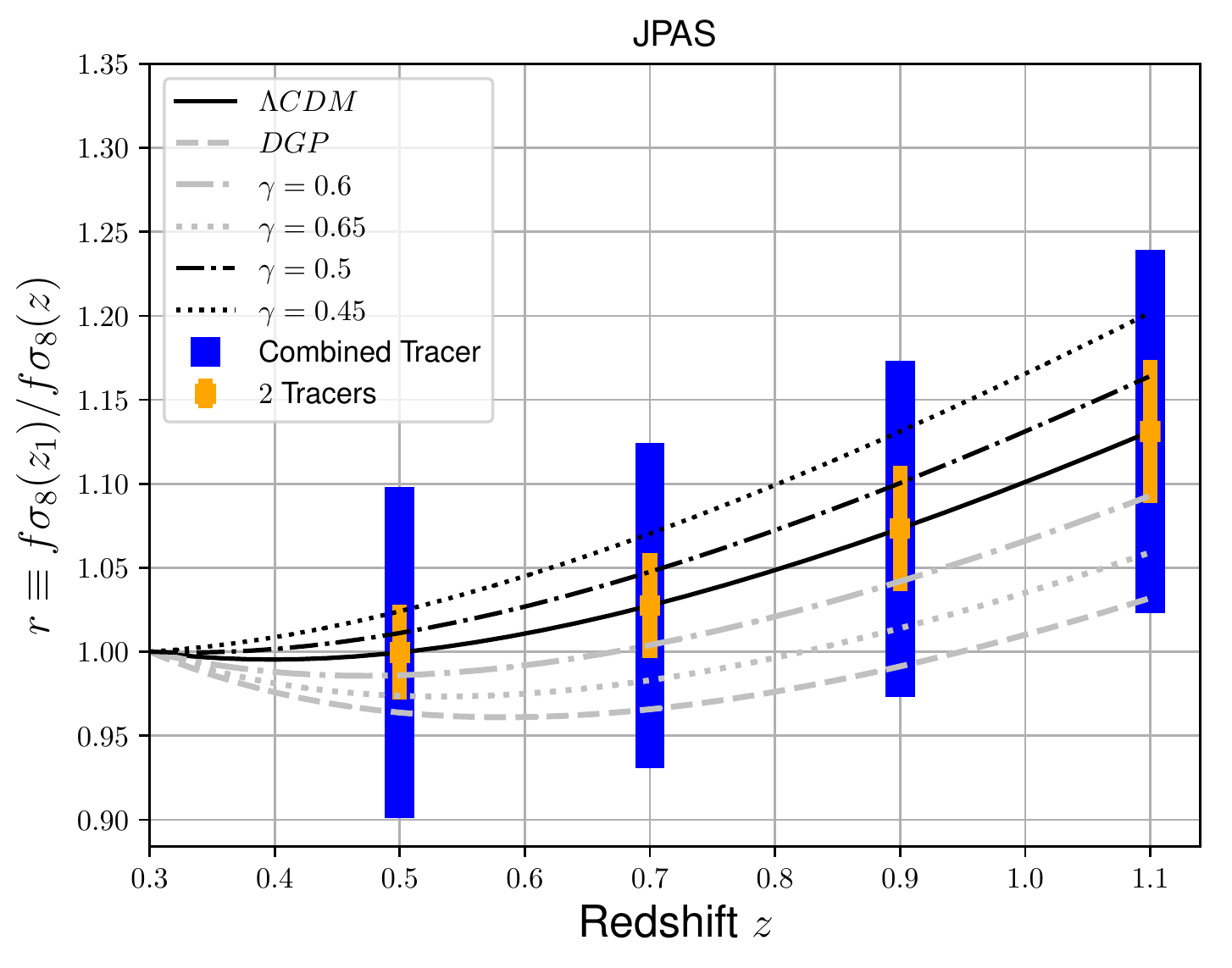}
    \includegraphics[scale = 0.5]{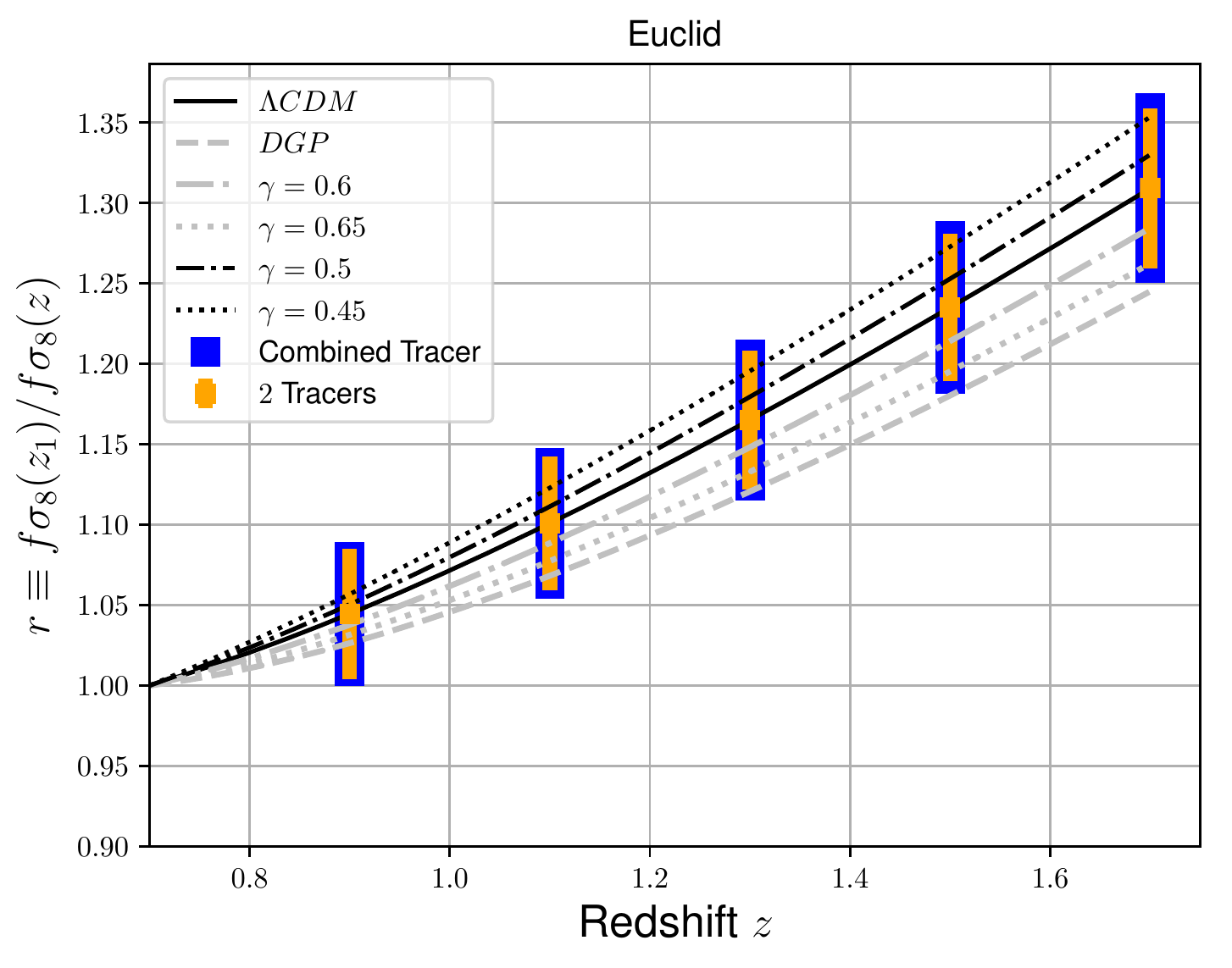}
    \includegraphics[scale = 0.5]{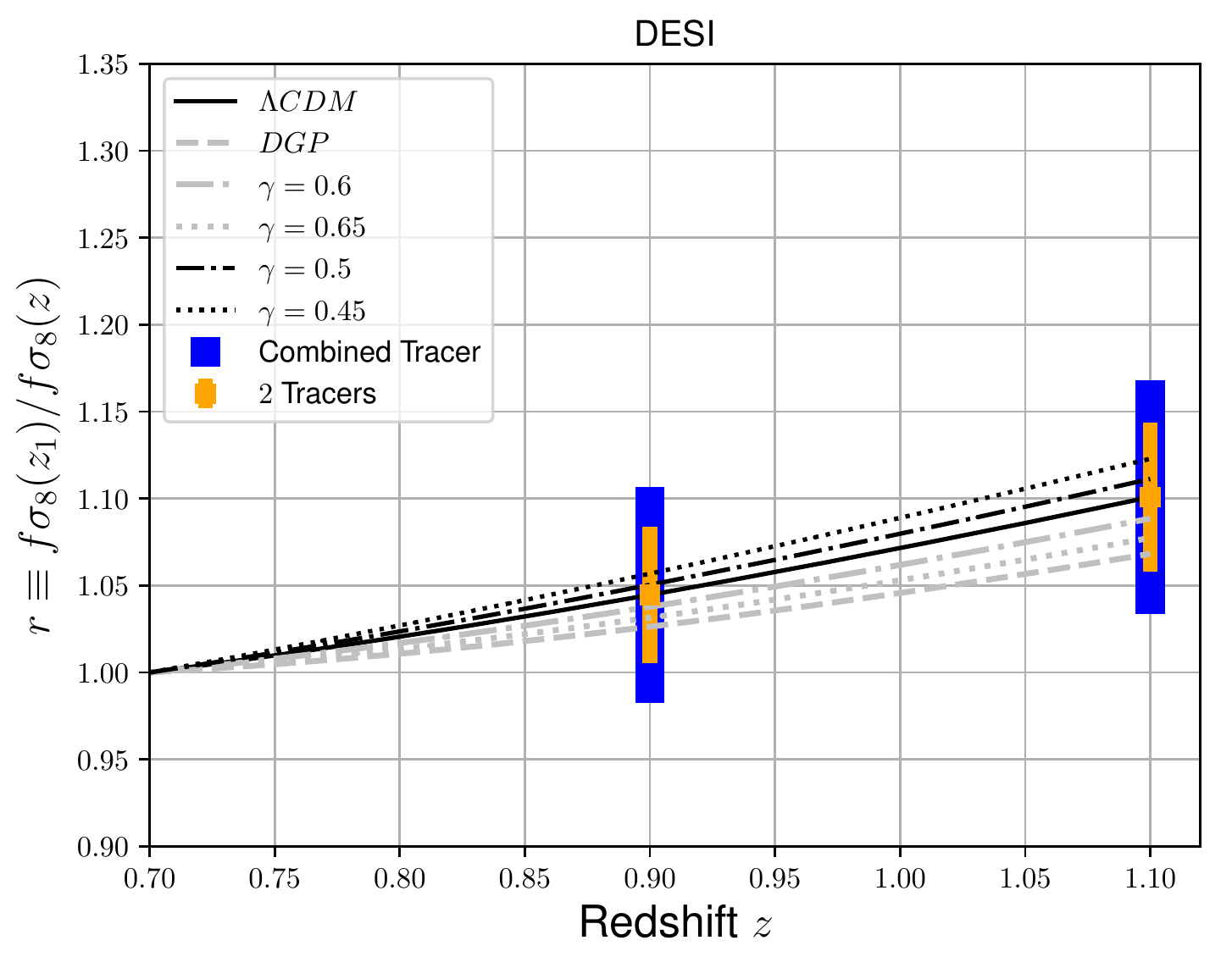}
    \caption{ 
    Constraints on $r$ for J-PAS (top left), Euclid (top right) and DESI (bottom middle) obtained using bins in the range $k =0.01 - 0.1[h $ Mpc$^{-1}]$ with the binning $\Delta k = 0.005 h $ Mpc$^{-1}$   . These results show how each survey is capable of distinguishing among models. The error bars in blue were drawn from the single-tracer Fisher matrix using a combination of two tracers, while orange error bars were drawn from the two-tracers Fisher matrix. For J-PAS, Euclid and DESI we use, respectively, the tracers (ELG, LRG), (ELG, QSO) and (ELG, LRG). The numerical values of error bars along with the fractional difference between one (combined) and two tracers are shown in Table \ref{table_errors}. }
    \label{surveys}
\end{figure}

In order to obtain constraints for the surveys, we need to calculate the Fisher matrix $F = \mathcal{V}\bar{F}$, where $\bar{F}$ is the Fisher matrix per unit of phase-space volume we used to obtain the previous results and $\mathcal{V}$ is the phase-space volume, which is of course highly dependent on the survey. For each survey, we calculate the phase-space volume after integrating over $\mu$ as:
\begin{eqnarray}
\mathcal{V} \equiv VV_{k} =  V \times \frac{1}{2} \frac{ 4 \pi k^2 \Delta_k}{(2\pi)^3}  = \frac{V^{2/3}k^{2}}{2\pi},
\label{volfactor}
\end{eqnarray}
where $\Delta_k=2\pi V^{-1/3}$, $V$ is the survey volume and $k$ is the scale at which the constraints were calculated. 
The volume for each survey is calculated assuming a redshift bin of $\Delta z = 0.2$ with central redshifts depending on the survey (see Table \ref{table_den}). Then the comoving volume of a spherical shell around each central redshift $z_{i}$ is $V = (4/3)\pi(\chi^{3}(z_{i}+0.1) - \chi^{3}(z_{i}-0.1))$, with
\begin{equation}
\chi(z) = \frac{c}{H_{0}}\int_{0}^{z}\frac{dz'}{E(z')}\nonumber,
\end{equation}
where $c/H_{0} = 2997.92 \ {\rm Mpc} \ h^{-1}$ and $E(z) = \sqrt{\Omega_{m}(1+z)^{3} + (1-\Omega_{m})}$ is the $\Lambda$CDM Hubble-Lema\^itre dimensionless parameter.
In order to obtain the comoving volume for each survey we multiply $V$ by the corresponding observed fraction of the sky, which are $f_{sky} = 0.206$, $0.363$ and $0.339$, respectively for J-PAS, Euclid and DESI. 

 We illustrate how to obtain constraints for the above surveys summing the contribution of various bins of $k$, namely the range $k =0.01 - 0.1[h $ Mpc$^{-1}]$ with the binning $\Delta k = 0.005 h $ Mpc$^{-1}$, in the case of one tracer. We adopt $k_{\rm min}=0.01 h$ Mpc$^{-1}$ as our reference scale because there is little gain in pushing $k$ to even lower values -- at least for the typical volumes spanned by the surveys we consider in this paper. On the other hand, the constraints are enhanced when we include higher values of $k$, since the phase-space volume grows with $\sim k^2$. 
 However, in that limit the amplitude of the power spectrum is also dropping quickly, which tends to wash out the difference between employing one tracer or two tracers. 
 Moreover, as we push the scales to $k \gtrsim 0.1 h$ Mpc$^{-1}$, the Kaiser approximation starts to break down due to non-linear effects, and even the parametrization of the FoG effect becomes less reliable. Hence, for the reasons above, here we adopt $k_{max} = 0.1 \, h$ Mpc$^{-1}$. Nevertheless, by doing the exercise of adopting $k_{max}=0.15 h$ Mpc$^{-1}$ we find an overall strengthening of the constraints of $30\% \sim 50\%$ for the two-tracers approach and of $50\% \sim 60\%$ for the one-tracer approach. We also make this code publicly available (link on the footnote of page 6) so the interested reader can test different configurations.

 In order to calculate the observable $r$ for different models, we use the well known parameterization for the growth rate \cite{peebles1980large, fry1985dynamical, lightman1990omega, wang1998cluster}:
\begin{eqnarray}
f = \Omega_{m}^{\gamma}(z).
\label{growth_par}
\end{eqnarray}
where $\gamma$ is the growth index, which is useful in order to parameterize deviations from General Relativity (GR), even though it has no direct physical meaning. The value this parameter assumes in the standard (GR based) model is $\gamma_{GR}  \simeq 0.5454$. 
Therefore, we can parameterize $r(z_{\bar{1}}, z, \gamma)$ as
\begin{eqnarray}
r(z_{\bar{1}}, z, \gamma) = \frac{f(z_{\bar{1}})\sigma_{8}(z_{\bar{1}})}{f(z)\sigma_{8}(z)} =\left[ \frac{\Omega_{m}(z_{\bar{1}})}{\Omega_{m}(z)}\right]^{\gamma}\exp \left(\int_{z_{\bar{1}}}^{z}\frac{\Omega_{m}^{\gamma}(z')}{1+z'}dz'\right).
\end{eqnarray}
where 
\begin{equation}
\Omega_{m}(z) = \frac{\Omega^{(0)}_{m}(1+z)^3}{E(z)}\,.
\end{equation}

In this exercise we fix one redshift slice at $z_{\bar{1}} = 0.3$ for J-PAS and at $z_{\bar{1}} = 0.7$ for Euclid and DESI, while the second redshift slice varies in a different range depending on the survey. 

Each survey will focus on a particular redshift range and hence will have different targets. Here we use two tracers for each survey and choose the two tracers in order to maximise the total signal. For instance, for DESI we could choose two tracers among ELGs, LRGs and QSOs. We choose to use ELGs and LRGs since these tracers are more abundant than QSOs in the redshift range we consider here for DESI. Similarly, for J-PAS and Euclid we use, respectively, (ELGs, LRGs) and (ELGs, QSOs). The densities of these tracers for each survey can be found in Table \ref{table_den}. Since there are no estimates for Euclid in literature for the density of quasars, in Table \ref{table_den} the densities of quasars are estimates for DESI. Therefore, we are assuming that Euclid will detect a similar density of quasars as DESI.

As  in the previous section, we compare a survey where the two tracers are treated individually to the case where they   are combined into a single one. For J-PAS and DESI we used the same fiducial bias values as were used by  Ref. \cite{resco2019j}, namely:
\begin{eqnarray}
 b(z)=\frac{b_{0}}{D(z)},
\end{eqnarray}
where $b_{0} = 0.84$ for ELGs and $b = 1.7$ for LRG. For Euclid we used bias of the form  $b(z) = \sqrt{1+z}$ for ELGs and $b(z) = 0.53 + 0.289(1+z)^2$ for QSOs \cite{laureijs2011euclid}. 

 \subsection{Constraints from future surveys}
 To obtain the constraints of this section, we start from the Fisher matrix per unit of phase-space,
 \begin{equation}
 \bar{F}[Y] = {\rm diag} \{ \bar{F}(z_{\bar{1}}, k_{1}, \mu), ..., \bar{F}(z_{\bar{1}}, k_{n}, \mu), \bar{F}(z_{\bar{2}}, k_{1}, \mu), ..., \bar{F}(z_{\bar{2}}, k_{n}, \mu)\} \; ,
 \end{equation}
 where
\begin{equation}
\bar{F}(z_{\bar{i}}, k_{\alpha}, \mu) =\left[  \frac{\mathcal{P}_{g}(z_{\bar{i}}, k_{\alpha}, \mu)}{1+ \mathcal{P}_{g}(z_{\bar{i}}, k_{\alpha}, \mu)} \right]^{2}
\end{equation}
is the Fisher matrix  in each $z$-shell and $k$-bin. 
The parameters of this Fisher matrix are the effective power in redshift-space, $\mathcal{P}_{g}(z, k, \mu)$, in each $z$-shell and $k$-bin:
\begin{equation}
Y = \{ \log \mathcal{P}(z_{\bar{1}}, k_{1}, \mu), ..., \log \mathcal{P}(z_{\bar{1}}, k_{n}, \mu), \log \mathcal{P}(z_{\bar{2}}, k_{1}, \mu), ..., \log \mathcal{P}(z_{\bar{2}}, k_{n}, \mu) \}.  
\end{equation}
Therefore, when considering two redshift slices, the set $Y$ has $2\times n_{k}$ parameters, where $n_{k}$ is the number of $k$-bins. In order to take into account the survey's volume we multiply this Fisher matrix by 
\begin{equation}
\mathcal{V} = {\rm diag}\{ \mathcal{V}(z_{\bar{1}}, k_{1}), ..., \mathcal{V}(z_{\bar{1}}, k_{n}),\mathcal{V}(z_{\bar{2}}, k_{1}), ..., \mathcal{V}(z_{\bar{2}}, k_{n})\}.
\end{equation}
We project the Fisher matrix $\mathcal{V}\bar{F}[Y]$ into the set
\begin{gather}
X = \{ \log r, \log P_{\bar{1}}(k_{1}), ..., \log P_{\bar{1}}(k_{n}),\log \beta_{\bar{1}}(k_{1}), ..., \log \beta_{\bar{1}}(k_{n}), \log \beta_{\bar{2}}(k_{1}), ..., \log \beta_{\bar{2}}(k_{n})\nonumber \\ 
,\log \sigma_{\bar{1} v}(k_{1}), ..., \log \sigma_{\bar{1} v}(k_{n}), \log \sigma_{\bar{2} v}(k_{1}), ..., \log \sigma_{\bar{2} v}(k_{n}) \}, 
\end{gather}
as in equation (\ref{project_single}). The combined quantities $\beta_{\bar{i}}$ and $\sigma_{\bar{i} v}$ are scale-dependent according to (\ref{fog_eff}) and (\ref{beta_eff}). Finally, the covariance is calculated averaging over $\mu$ as in equation (\ref{mu_ave}) and inverting the Fisher matrix for the set $X$.  As pointed out in the previous section, we add conservative priors of $50\%$ on the dispersion velocities. For two tracers, the starting point is a block diagonal matrix, where the block diagonals are the two-tracers Fisher matrices (\ref{Fisher_2t}) for each $z$-shell and $k$-bin, followed by the same procedure.

We report in Fig. (\ref{surveys}) how J-PAS, Euclid, and DESI surveys are capable of distinguishing among models of gravity, contrasting the one (combined) and two tracers cases. In this figure we show predictions for $r$ using generic modified gravity models with $\gamma = 0.45,$ $0.5,$ $0.6$ and $0.65$, together with the prediction from the DGP (Dvali-Gabadadze-Porrati) model \cite{dvali20004d, dvali2001gravity}  using the $\gamma_{DGP}$ parameterization of Ref. \cite{linder2007parameterized}. 
Figure (\ref{surveys}) shows error bars for the model-independent quantity $r$, overimposed to  the prediction from the generic parameterization of Eq. (\ref{growth_par}). It is important to remark that our constraints on $r$ apply independently in any redshift shell.

For J-PAS the difference between one combined tracer and two tracers is more significant. For instance, the two-tracers constraints for J-PAS can distinguish $\gamma = 0.65$ and $\gamma = 0.45$ from $\Lambda$CDM, but the combined tracers constraints can not for any $z-$shell. This advantage appears due to the larger value of $P_{\bar{2}}$ at the second redshift slice in the J-PAS case (see Table \ref{table_den}). The larger signal arises because J-PAS  is a photometric survey and hence will detect a larger number of objects compared to Euclid and DESI, which are spectroscopic surveys. Furthermore, since QSOs are sparse tracers of the LSS, they have much less signal compared with ELGs or LRGs, hence the error bars for Euclid in the two cases (combined and two tracers) are even more similar. As it is clear from previous results, the difference between one and two tracers depends mainly on the SNR. 

Figure (\ref{surveys}) shows that the way in which the SNR is distributed among the species is also relevant to determine the effectiveness of the two tracers approach.  Moreover, the J-PAS survey is more capable of distinguishing among models with different $\gamma$s, although the error bars are similar to those from Euclid and DESI. This is due to the fact that at low values of redshift, $f\sigma_{8}$, and as consequence $r$, is much more sensitive to $\gamma$ than it is at high values of redshift.

In Table \ref{table_errors_beta} we show, for each survey, the relative marginalized errors on $\beta_{\bar{2}}$, $\sigma_{\beta_{\bar{2}}}$, in the case where these errors are drawn directly from the single-tracer Fisher matrix and in the case where we propagate the errors $\sigma_{\beta_{\bar{2}1}}$, $\sigma_{\beta_{\bar{2}2}}$ and $\sigma_{\beta_{\bar{2}1}\beta_{\bar{2}2}}$ from the two-tracers Fisher matrix using the usual uncertainty propagating formula:
\begin{equation}
    (\sigma^{2t}_{\beta_{\bar{2}}})^{2}=\frac{\beta_{\bar{2}2}^{2} \sigma_{\beta_{\bar{2}1}}^{2}+q_{\bar{2}}^{2} \beta_{\bar{2}1}^{2} \sigma_{\beta_{\bar{2}2}}^{2}+2 q_{\bar{2}} \beta_{\bar{2}1} \beta_{\bar{2}2} \sigma_{\beta_{\bar{2}1} \beta_{\bar{2}2}}}{\left(\beta_{\bar{2}2}+\beta_{\bar{2}1} q_{\bar{2}}\right)^{2}}.
    \label{propagating}
\end{equation}
To obtain the above formula we assume that the combined $\beta_{\bar{2}}$ is given by
\begin{equation}
    \beta_{\bar{2}} =\frac{(1+q_{\bar{2}}) \beta_{\bar{2}1} \beta_{\bar{2}2}}{\beta_{\bar{2}2}+ q_{\bar{2}} \beta_{1\bar{2}}}.
\end{equation}
Note that here we are assuming that the combined $\beta_{\bar{2}}$ is $k$-independent, following our assumption in this section that $P_{\bar{1}}$ is the only $k$-dependent parameter of the single-tracer Fisher matrix. Looking at equation (\ref{beta_eff}), we can make the combined $\beta$ a $k$-independent quantity by setting $\sigma_{1} = \sigma_{2}$.

The comparison between the relative marginalized error $\sigma^{1t}_{\beta_{\bar{2}}}$ drawn directly from the single-tracer Fisher matrix and the propagated $\sigma^{2t}_{\beta_{\bar{2}}}$ (see Table \ref{table_errors_beta}) shows that the relative gain of two-tracers approach is always quite important for $\beta$. In Table \ref{table_errors_beta} we also show the constraints on $\sigma_{v}$ for the single-tracer approach. 

The code used to obtain the results of this section is publicly available (the link on the footnote is in Section II.C), so the interested reader can explore these results for other configurations ($k$-bins, volumes, etc.).

\renewcommand{\arraystretch}{1.3}
\begin{center}
\begin{table}
\centering
\begin{tabular}{ ccccccccc }   
\begin{tabular}{ P{0.6cm} | P{1cm} P{1cm} P{2cm} P{1cm} | P{1cm} P{1cm} P{2cm} P{1cm} | P{1cm} P{1cm}  P{2cm} P{1cm} } 
\hline
    \hline
    & \multicolumn{4}{c|}{J-PAS} & 
    \multicolumn{4}{c|}{Euclid} & 
    \multicolumn{4}{c}{DESI} \\
    \hline
    $z$ & $\bar{n}^{ELG}$ & $\bar{n}^{LRG}$ & $V(\times10^{9}Mpc^{3})$ & $P_{\bar{2}}$ &  $\bar{n}^{ELG}$ & $\bar{n}^{QSO}$ & $V(\times10^{9}Mpc^{3})$ & $P_{\bar{2}}$ &  $\bar{n}^{ELG}$ & $\bar{n}^{LRG}$ & $V(\times10^{9}Mpc^{3})$ & $P_{\bar{2}}$  \\
    \hline
    0.5   &    1181.1  & 156.3 &  6.63 & 253.43 &    --   &    --  &  --  &   --  &  --  &   --  &  --  &  --  \\
    0.7   &     502.1  &  68.8 & 10.3  & 108.9  &    --   &    --  &  --  &   --  &  --  &   --  &  --  &  --  \\
    0.9   &     138.0  &  12.0 & 13.5  &  26.54 &  206.6  &  2.6   & 23.8 & 31.16 & 81.9 & 19.1  & 22.3 & 19.59\\
    1.1   &      41.2  &   0.9 & 16.2  &   6.64 &  161.5  &  2.555 & 28.5 & 22.76 & 47.7 &  1.18 & 26.6 &  6.99\\
    1.3   &       --   &   --  &   --  &   --   &  121.25 &  2.5   & 32.1 & 16.07 &  --  &  --   &  --  &   -- \\
    1.5   &       --   &   --  &   --  &   --   &   81.75 &  2.4   & 34.8 & 10.32 &  --  &  --   &  --  &   -- \\
    1.7   &       --   &   --  &   --  &   --   &   50.25 &  2.3   & 36.6 &  6.18 &  --  &  --   &  --  &   -- \\
  \hline\hline\end{tabular}  

\end{tabular}
\caption{Left: Galaxy Densities (ELG and LRG), volumes and SNR at the second redshift slice, $P_{\bar{2}}$, for
J-PAS at each redshift bin. Middle: Galaxy densities (ELG and QSO), volumes and  $P_{\bar{2}}$ for Euclid at each redshift bin. Right: Galaxy densities (ELG and LRG), volumes and  $P_{\bar{2}}$ for
DESI at each redshift bin. Galaxy densities in units of $10^{-5} \ h^{3}$ Mpc$^{-3}$. Here, $P_{\bar{2}}$ is evaluated at $k = 0.01 h$ Mpc$^{-1}$. The values of number densities was extracted from references \cite{costa2019j} and \cite{aparicio2020j}. The number densities for quasars in the Euclid survey were extrapolated from the values for the DESI survey. }
\label{table_den}
\end{table}
\end{center}

\begin{center}
\begin{table}
\centering
\begin{tabular}{ ccccccccc }   
\begin{tabular}{ P{0.8cm} |  P{1cm}  P{1cm} P{1cm} |P{1cm} P{1cm} P{1cm}|P{1cm} P{1cm} P{1cm}} 
\hline
    \hline
    & &  J-PAS & & &  Euclid & & & DESI &  \\ 
    \hline
    $z$ &  $\sigma^{1t}_{r}$ &  $\sigma^{2t}_{r}$ &  $\Delta_{r}$ & $\sigma^{1t}_{r}$ &  $\sigma^{2t}_{r}$ &  $\Delta_{r}$ & $\sigma^{1t}_{r}$ &  $\sigma^{2t}_{r}$ &  $\Delta_{r}$ \\
    \hline
    0.5   &     0.1   & 0.028 & 2.16 &  --   &   --  &   --  &   --  &   --  & --  \\
    0.7   &     0.094 & 0.03  & 1.37 &  --   &   --  &   --  &   --  &   --  & --  \\
    0.9   &     0.093 & 0.035 & 0.77 & 0.043 & 0.039 & 0.1   & 0.059 & 0.038 & 0.58\\
    1.1   &     0.095 & 0.037 & 0.66 & 0.042 & 0.037 & 0.13  & 0.060 & 0.039 & 0.57\\
    1.3   &      --   &   --  &  --  & 0.043 & 0.037 & 0.15  &   --  &   --  & --  \\
    1.5   &      --   &   --  &  --  & 0.043 & 0.037 & 0.17  &   --  &   --  & --  \\
    1.7   &      --   &   --  &  --  & 0.045 & 0.038 & 0.18  &   --  &   --  & --  \\
  \hline\hline\end{tabular}  

\end{tabular}
\caption{ Numerical relative marginalized errors for the one (combined) tracer and two tracers approach, $\sigma^{1t}_{r}$ and $\sigma^{2t}_{r}$, and the fractional gain of the two tracers errors, $\Delta_{r} \equiv \sigma^{1t}_{r}/\sigma^{2t}_{r}-1 $, for each survey. These are the numerical values of errors shown in Figure (\ref{surveys}).}
\label{table_errors}
\end{table}
\end{center}

\begin{center}
\begin{table}
\centering
\begin{tabular}{ ccccccccc }   
\begin{tabular}{ P{0.8cm} |  P{1.cm}  P{1.cm} P{1cm} P{1cm} |P{1.cm} P{1.cm} P{1cm} P{1cm}|P{1.cm} P{1.cm} P{1cm} P{1cm}} 
\hline
    \hline
    & \multicolumn{4}{c|}{J-PAS} & 
    \multicolumn{4}{c|}{Euclid} & 
    \multicolumn{4}{c}{DESI} \\
    \hline
    $z$ &  $\sigma^{1t}_{\beta_{\bar{2}}}$ &  $\sigma^{2t}_{\beta_{\bar{2}}}$ &  $\Delta_{\beta_{\bar{2}}}$ & $\sigma^{1t}_{\sigma_{v}}$&  $\sigma^{1t}_{\beta_{\bar{2}}}$ &  $\sigma^{2t}_{\beta_{\bar{2}}}$ &  $\Delta_{\beta_{\bar{2}}}$ & $\sigma^{1t}_{\sigma_{v}}$ & $\sigma^{1t}_{\beta_{\bar{2}}}$ &  $\sigma^{2t}_{\beta_{\bar{2}}}$ &  $\Delta_{\beta_{\bar{2}}}$ & $\sigma^{1t}_{\sigma_{v}}$ \\
    \hline
    0.5   &     0.066 & 0.021 & 2.16 &  0.4  &  --   &   --  &   -- & --     &   --  &   --  & --   & --\\
    0.7   &     0.058 & 0.024 & 1.37 &  0.4  &  --   &   --  &   -- & --     &   --  &   --  & --   & -- \\
    0.9   &     0.055 & 0.031 & 0.77 &  0.4  & 0.033 & 0.029 & 0.13 &  0.38  & 0.046 & 0.028 & 0.63 & 0.39 \\
    1.1   &     0.059 & 0.035 & 0.66 &  0.4  & 0.032 & 0.027 & 0.18 &  0.38  & 0.048 & 0.031 & 0.54 & 0.39 \\
    1.3   &      --   &   --  &  --  &  --   & 0.032 & 0.026 & 0.23 &  0.38  &   --  &   --  & --   &  -- \\
    1.5   &      --   &   --  &  --  &  --   & 0.033 & 0.026 & 0.27 &  0.38  &   --  &   --  & --   & -- \\
    1.7   &      --   &   --  &  --  &  --   & 0.035 & 0.027 & 0.27 &  0.38  &   --  &   --  & --   & -- \\
  \hline\hline\end{tabular}  

\end{tabular}
\caption{  Numerical relative errors for the one (combined) tracer and two tracers approach, $\sigma^{1t}_{\beta_{\bar{2}}}$ and $\sigma^{2t}_{\beta_{\bar{2}}}$, and the fractional gain of the two tracers errors, $\Delta_{\beta_{\bar{2}}} \equiv \sigma^{1t}_{\beta_{\bar{2}}}/\sigma^{2t}_{\beta_{\bar{2}}}-1 $, as well as the constraints for the combined pairwise velocity dispersion, $\sigma^{1t}_{v}$, for each survey. The two tracers constraints for $\beta_{\bar{2}}$ are obtained by propagating from $\sigma_{\beta_{\bar{2}1}}$, $\sigma_{\beta_{\bar{2}2}}$ and $\sigma_{\beta_{\bar{2}1}}$ to $\sigma_{\beta_{\bar{2}}}$ through equation (\ref{propagating}).}
\label{table_errors_beta}
\end{table}
\end{center}

\section{Discussion and Conclusions}

In this work we have derived model-independent  constraints on $f\sigma_{8}(z)$ by combining two redshift bins through the observable $r\equiv f\sigma_{8}(z_{\bar{1}})/f\sigma_{8}(z_{\bar{2}})$. 
Here, model-independent means that we do not need to assume any specific cosmological model, e.g. $\Lambda$CDM, in order to derive the constraints: we leave the power spectrum, as well as any other quantities like the redshift distortion  $\beta$, as  parameters free to vary in $k$ and $z$.

We had two goals in mind. First, to determine what is the complete set of quantities that can be estimated from the linear power spectrum (with the mild non-linear Fingers-of-God corrections) without assuming a cosmological model.
Secondly, to 
obtain  constraints in particular on the statistics $r(k,z_1,z_2)$ for various future surveys and
 to assess the advantage of two tracers with respect to the standard single tracer surveys.

Concerning the first goal, we found that there are eight statistics  (see Eq. \ref{X2t}) that can be obtained combining two redshift bins and two tracers. This is the maximal model-independent set in the clustering linear regime. Every other combination is either degenerate with this set, or is not a model-independent quantity. Additional quantities can  be introduced using lensing and peculiar velocities, and will be studied elsewhere.

Concerning the second point,
we contrasted two situations: (i) a survey with two distinct tracers of large-scale structure, and (ii) the same survey but  with a single-tracer which is a combination of the two distinct tracers. 
The quantification of the difference between the two cases is important in order to set observational strategies in near future surveys. 

In section \ref{results} we explore how the relative marginalised error $\sigma_{r}$ depends on the SNR ($P_{\bar{1}}$) and the redshift-space distortion parameter $\beta_{\bar{1}1}$ of the first redshift slice. 
As found in \cite{2019JCAP...06..030A} for the redshift-space parameter $\beta$, the difference between one and two-tracers constraints strongly depends on the SNR (expressed in terms of the clustering strength at a given redshift slice, $P_{\bar{z}}$), as well as on the difference between $\beta_{\bar{1}1}$ and $\beta_{\bar{1}2}$. 
For SNR larger than 1, the two-tracers approach is increasingly more advantageous.

In section \ref{surveys_section} we show how the observational strategies of three near future surveys perform for one and two-tracers.
Both DESI and J-PAS, which will observe mostly ELGs and LRGs, benefit greatly (gains $\gtrsim 50$ \%) from having at least two distinct types of tracers over some redshift range. Euclid, which combines ELGs and quasars over the same redshift range, presents a more modest gain in $r$, of $ 10\% \sim 20\%$.
We find that $r$ can be measured to within 4-10\% for one tracer, and to 3-6\% with two tracers.

In this work we provide further evidence that the two-tracers approach is always more advantageous, in particular when measuring physical observables in a model-independent way. 
We have also shown that the gains accrued by distinguishing between the tracers depend not only on the signal available, but also on how that signal is distributed between the two tracers. The main conclusion is that, when two tracers have sufficiently distinct biases (or, equivalently, distinct RSD parameters $\beta$), and their SNR are not both $\ll 1$, then it is significantly advantageous to keep them as separated tracers for the sake of extracting physical parameters. 
In other words, theoretically, there is no downside in splitting the sample of tracers. 
In practice, of course, it may be difficult to measure the clustering of an extremely sparse tracer, although this has been proved possible both in $N$-body simulations \cite{Dorta20} as well as in realistic surveys \cite{Gabi19}.

\section*{Acknowledgments}

L. R. A. would like to thank CNPq and FAPESP (grant 2018/04683-9) for support; L.A. acknowledges support from DAAD to attend the II Workshop on Current Challenges in Cosmology, Bogotá (Colombia),   where this work was started; R.B. would like to thank CAPES for financial support.

\bibliographystyle{plain}
\bibliography{scaling_bib,biblist,refs}

\begin{thebibliography}{10}

\bibitem{abramo2012full}
L~Raul Abramo.
\newblock The full fisher matrix for galaxy surveys.
\newblock {\em Mon. Not. Roy. Astron. Soc.}, 420(3):2042--2057, 2012.

\bibitem{2019JCAP...06..030A}
L.~Raul {Abramo} and Luca {Amendola}.
\newblock {Fisher matrix for multiple tracers: model independent constraints on
  the redshift distortion parameter}.
\newblock {\em JCAP}, 2019(6):030, Jun 2019.

\bibitem{abramo2013multitracer}
L~Raul Abramo and Katie~E Leonard.
\newblock Why multitracer surveys beat cosmic variance.
\newblock {\em Mon. Not. Roy. Astron. Soc.}, 432(1):318--326, 2013.

\bibitem{aghamousa2016desi}
Amir Aghamousa, Jessica Aguilar, Steve Ahlen, Shadab Alam, Lori~E Allen,
  Carlos~Allende Prieto, James Annis, Stephen Bailey, Christophe Balland, Otger
  Ballester, et~al.
\newblock The desi experiment part i: Science, targeting, and survey design.
\newblock {\em arXiv preprint arXiv:1611.00036}, 2016.

\bibitem{aghanim2018planck}
N~Aghanim, Yashar Akrami, M~Ashdown, J~Aumont, C~Baccigalupi, M~Ballardini,
  AJ~Banday, RB~Barreiro, N~Bartolo, S~Basak, et~al.
\newblock Planck 2018 results. vi. cosmological parameters.
\newblock {\em arXiv preprint arXiv:1807.06209}, 2018.

\bibitem{alcock1979evolution}
Charles Alcock and Bohdan Paczy{\'n}ski.
\newblock An evolution free test for non-zero cosmological constant.
\newblock {\em Nature}, 281(5730):358--359, 1979.

\bibitem{amendola2018cosmology}
Luca Amendola, Stephen Appleby, Anastasios Avgoustidis, David Bacon, Tessa
  Baker, Marco Baldi, Nicola Bartolo, Alain Blanchard, Camille Bonvin, Stefano
  Borgani, et~al.
\newblock Cosmology and fundamental physics with the euclid satellite.
\newblock {\em Living reviews in relativity}, 21(1):2, 2018.

\bibitem{Amendola:2019lvy}
Luca Amendola and Miguel Quartin.
\newblock {Measuring the Hubble function with standard candle clustering}.
\newblock 2019.

\bibitem{aparicio2020j}
Miguel Aparicio~Resco, Antonio~L Maroto, Jailson~S Alcaniz, L~Raul Abramo,
  C~Hern{\'a}ndez-Monteagudo, N~Ben{\'\i}tez, S~Carneiro, AJ~Cenarro,
  D~Crist{\'o}bal-Hornillos, RA~Dupke, et~al.
\newblock J-pas: forecasts on dark energy and modified gravity theories.
\newblock {\em Monthly Notices of the Royal Astronomical Society},
  493(3):3616--3631, 2020.

\bibitem{benitez2014j}
N~Benitez, R~Dupke, M~Moles, L~Sodre, J~Cenarro, A~Marin-Franch, K~Taylor,
  D~Cristobal, A~Fernandez-Soto, C~Mendes de~Oliveira, et~al.
\newblock J-pas: the javalambre-physics of the accelerated universe
  astrophysical survey.
\newblock {\em arXiv preprint arXiv:1403.5237}, 2014.

\bibitem{2013MNRAS.436..854B}
S.~{Benitez-Herrera}, E.~E.~O. {Ishida}, M.~{Maturi}, W.~{Hillebrandt},
  M.~{Bartelmann}, and F.~{R{\"o}pke}.
\newblock {Cosmological parameter estimation from SN Ia data: a
  model-independent approach}.
\newblock {\em MNRAS}, 436(1):854--858, November 2013.

\bibitem{beutler20126df}
Florian Beutler, Chris Blake, Matthew Colless, D~Heath Jones, Lister
  Staveley-Smith, Gregory~B Poole, Lachlan Campbell, Quentin Parker, Will
  Saunders, and Fred Watson.
\newblock The 6df galaxy survey: z= 0 measurements of the growth rate and
  $\sigma$8.
\newblock {\em Monthly Notices of the Royal Astronomical Society},
  423(4):3430--3444, 2012.

\bibitem{2018JCAP...05..008C}
Salvatore {Capozziello}, Rocco {D'Agostino}, and Orlando {Luongo}.
\newblock {Rational approximations of f(R) cosmography through Pad'e
  polynomials}.
\newblock {\em JCAP}, 2018(5):008, May 2018.

\bibitem{carron2015information}
Julien Carron, Melody Wolk, and Istv{\'a}n Szapudi.
\newblock On the information content of the matter power spectrum.
\newblock {\em Mon. Not. Roy. Astron. Soc.}, 453(1):450--455, 2015.

\bibitem{costa2019j}
AA~Costa, RJF Marcondes, RG~Landim, E~Abdalla, LR~Abramo, HS~Xavier, AA~Orsi,
  N~Chandrachani Devi, AJ~Cenarro, D~Crist{\'o}bal-Hornillos, et~al.
\newblock J-pas: forecasts on interacting dark energy from baryon acoustic
  oscillations and redshift-space distortions.
\newblock {\em Monthly Notices of the Royal Astronomical Society},
  488(1):78--88, 2019.

\bibitem{de2013vimos}
S~De~La~Torre, L~Guzzo, JA~Peacock, E~Branchini, A~Iovino, BR~Granett, U~Abbas,
  C~Adami, S~Arnouts, Julien Bel, et~al.
\newblock The vimos public extragalactic redshift survey (vipers)-galaxy
  clustering and redshift-space distortions at z= 0.8 in the first data
  release.
\newblock {\em Astronomy \& Astrophysics}, 557:A54, 2013.

\bibitem{dvali2001gravity}
Gia Dvali and Gregory Gabadadze.
\newblock Gravity on a brane in infinite-volume extra space.
\newblock {\em Physical Review D}, 63(6):065007, 2001.

\bibitem{dvali20004d}
Gia Dvali, Gregory Gabadadze, and Massimo Porrati.
\newblock 4d gravity on a brane in 5d minkowski space.
\newblock {\em Physics Letters B}, 485(1-3):208--214, 2000.

\bibitem{2019JCAP...12..028F}
Jos{\'e} {Fonseca}, Jan-Albert {Viljoen}, and Roy {Maartens}.
\newblock {Constraints on the growth rate using the observed galaxy power
  spectrum}.
\newblock {\em JCAP}, 2019(12):028, December 2019.

\bibitem{Fonseca:2015laa}
José Fonseca, Stefano Camera, Mário Santos, and Roy Maartens.
\newblock {Hunting down horizon-scale effects with multi-wavelength surveys}.
\newblock {\em Astrophys. J.}, 812(2):L22, 2015.

\bibitem{fry1985dynamical}
James~N Fry.
\newblock Dynamical measures of density in exotic cosmologies.
\newblock {\em Physics Letters B}, 158(3):211--214, 1985.

\bibitem{hamilton1998linear}
AJS Hamilton.
\newblock Linear redshift distortions: a review.
\newblock In {\em The evolving universe}, pages 185--275. Springer, 1998.

\bibitem{howlett2015clustering}
Cullan Howlett, Ashley~J Ross, Lado Samushia, Will~J Percival, and Marc Manera.
\newblock The clustering of the sdss main galaxy sample--ii. mock galaxy
  catalogues and a measurement of the growth of structure from redshift space
  distortions at z= 0.15.
\newblock {\em Monthly Notices of the Royal Astronomical Society},
  449(1):848--866, 2015.

\bibitem{2020arXiv200110887K}
Hanwool {Koo}, Arman {Shafieloo}, Ryan~E. {Keeley}, and Benjamin {L'Huillier}.
\newblock {Model-independent constraints on Type Ia supernova light-curve
  hyper-parameters and reconstructions of the expansion history of the
  Universe}.
\newblock {\em arXiv e-prints}, page arXiv:2001.10887, January 2020.

\bibitem{laureijs2011euclid}
Rene Laureijs, J~Amiaux, S~Arduini, J-L Augueres, J~Brinchmann, R~Cole,
  M~Cropper, C~Dabin, L~Duvet, A~Ealet, et~al.
\newblock Euclid definition study report.
\newblock {\em arXiv preprint arXiv:1110.3193}, 2011.

\bibitem{2019ApJ...887...36L}
Shi-Yu {Li}, Yun-Long {Li}, Tong-Jie {Zhang}, and Tingting {Zhang}.
\newblock {Model-independent Determination of Cosmic Curvature Based on the
  Pad{\'e} Approximation}.
\newblock {\em APJ}, 887(1):36, December 2019.

\bibitem{lightman1990omega}
Alan~P Lightman and Paul~L Schechter.
\newblock The omega dependence of peculiar velocities induced by spherical
  density perturbations.
\newblock {\em The Astrophysical Journal Supplement Series}, 74:831, 1990.

\bibitem{linder2007parameterized}
Eric~V Linder and Robert~N Cahn.
\newblock Parameterized beyond-einstein growth.
\newblock {\em Astroparticle Physics}, 28(4-5):481--488, 2007.

\bibitem{McDonald:2009}
Patrick McDonald and Uros Seljak.
\newblock How to measure redshift-space distortions without sample variance.
\newblock {\em JCAP}, 0910:007, 2009.

\bibitem{Dorta20}
Antonio~D. {Montero-Dorta}, L.~Raul {Abramo}, Benjamin~R. {Granett}, Sylvain
  {de la Torre}, and Luigi {Guzzo}.
\newblock {The Multi-Tracer Optimal Estimator applied to VIPERS}.
\newblock {\em Mon. Not. Roy. Astron. Soc.}, 493(4):5257--5272, April 2020.

\bibitem{2009JCAP...01..044M}
Edvard {M{\"o}rtsell} and Chris {Clarkson}.
\newblock {Model independent constraints on the cosmological expansion rate}.
\newblock {\em JCAP}, 2009(1):044, January 2009.

\bibitem{peacock1994reconstructing}
JA~Peacock and SJ~Dodds.
\newblock Reconstructing the linear power spectrum of cosmological mass
  fluctuations.
\newblock {\em Mon. Not. Roy. Astron. Soc.}, 267(4):1020--1034, 1994.

\bibitem{peebles1980large}
Phillip James~Edwin Peebles.
\newblock {\em The large-scale structure of the universe}.
\newblock Princeton university press, 1980.

\bibitem{perico2019cosmic}
Eder~LD Perico, Rodrigo Voivodic, Marcos Lima, and David~F Mota.
\newblock Cosmic voids in modified gravity scenarios.
\newblock {\em Astronomy \& Astrophysics}, 632:A52, 2019.

\bibitem{reid2012clustering}
Beth~A Reid, Lado Samushia, Martin White, Will~J Percival, Marc Manera, Nikhil
  Padmanabhan, Ashley~J Ross, Ariel~G S{\'a}nchez, Stephen Bailey, Dmitry
  Bizyaev, et~al.
\newblock The clustering of galaxies in the sdss-iii baryon oscillation
  spectroscopic survey: measurements of the growth of structure and expansion
  rate at z= 0.57 from anisotropic clustering.
\newblock {\em Monthly Notices of the Royal Astronomical Society},
  426(4):2719--2737, 2012.

\bibitem{resco2019j}
Miguel~Aparicio Resco, Antonio~L Maroto, Jailson~S Alcaniz, L~Raul Abramo,
  C~Hern{\'a}ndez-Monteagudo, N~Ben{\'\i}tez, S~Carneiro, AJ~Cenarro,
  D~Crist{\'o}bal-Hornillos, RA~Dupke, et~al.
\newblock J-pas: forecasts on dark energy and modified gravity theories.
\newblock {\em arXiv preprint arXiv:1910.02694}, 2019.

\bibitem{samushia2014clustering}
Lado Samushia, Beth~A Reid, Martin White, Will~J Percival, Antonio~J Cuesta,
  Gong-Bo Zhao, Ashley~J Ross, Marc Manera, Eric Aubourg, Florian Beutler,
  et~al.
\newblock The clustering of galaxies in the sdss-iii baryon oscillation
  spectroscopic survey: measuring growth rate and geometry with anisotropic
  clustering.
\newblock {\em Monthly Notices of the Royal Astronomical Society},
  439(4):3504--3519, 2014.

\bibitem{Gabi19}
Gabriela {Sato-Polito}, Antonio~D. {Montero-Dorta}, L.~Raul {Abramo}, Francisco
  {Prada}, and Anatoly {Klypin}.
\newblock {The dependence of halo bias on age, concentration, and spin}.
\newblock {\em Mon. Not. Roy. Astron. Soc.}, 487(2):1570--1579, August 2019.

\bibitem{Seljak:2009}
Uros Seljak.
\newblock Measuring primordial non-gaussianity without cosmic variance.
\newblock {\em Phys.Rev.Lett.}, 102:021302, 2009.

\bibitem{song2009reconstructing}
Yong-Seon Song and Will~J Percival.
\newblock Reconstructing the history of structure formation using redshift
  distortions.
\newblock {\em Journal of Cosmology and Astroparticle Physics}, 2009(10):004,
  2009.

\bibitem{Taddei:2014wqa}
Laura Taddei and Luca Amendola.
\newblock {A cosmological exclusion plot: Towards model-independent constraints
  on modified gravity from current and future growth rate data}.
\newblock {\em JCAP}, 02:001, 2015.

\bibitem{Taddei:2016iku}
Laura Taddei, Matteo Martinelli, and Luca Amendola.
\newblock {Model-independent constraints on modified gravity from current data
  and from the Euclid and SKA future surveys}.
\newblock {\em JCAP}, 12:032, 2016.

\bibitem{tegmark1997measuring}
Max Tegmark.
\newblock Measuring cosmological parameters with galaxy surveys.
\newblock {\em Physical Review Letters}, 79(20):3806, 1997.

\bibitem{Thomas:2011sf}
Daniel~B. Thomas and Carlo~R. Contaldi.
\newblock {Testing model independent modified gravity with future large scale
  surveys}.
\newblock {\em JCAP}, 12:013, 2011.

\bibitem{vargas2019unraveling}
Mariana Vargas-Magana, David~D Brooks, Michael~M Levi, and Gregory~G Tarle.
\newblock Unraveling the universe with desi.
\newblock {\em arXiv preprint arXiv:1901.01581}, 2019.

\bibitem{wang1998cluster}
Limin Wang and Paul~J Steinhardt.
\newblock Cluster abundance constraints for cosmological models with a
  time-varying, spatially inhomogeneous energy component with negative
  pressure.
\newblock {\em The Astrophysical Journal}, 508(2):483, 1998.

\bibitem{white2009forecasting}
Martin White, Yong-Seon Song, and Will~J Percival.
\newblock Forecasting cosmological constraints from redshift surveys.
\newblock {\em Mon. Not. Roy. Astron. Soc.}, 397(3):1348--1354, 2009.

\end{thebibliography}

\end{document}